\documentclass[prl,twocolumn,superscriptaddress,showpacs,nobalancelastpage]{revtex4}

\usepackage{stmaryrd}

\usepackage[pdftex]{graphicx}
\usepackage{float}

\usepackage{amssymb,amsmath}

\begin{document}

\title{Revealing topological Dirac fermions at the surface of strained HgTe thin films via Quantum Hall transport spectroscopy.}

\author{C. Thomas}
\affiliation{Univ. Grenoble Alpes, CEA, LETI, MINATEC campus, F38054 Grenoble, France.}

\author{O. Crauste}
\affiliation{CNRS, Inst NEEL, F-38042 Grenoble, France.}
\affiliation{Univ. Grenoble Alpes, Inst NEEL, F-38042 Grenoble, France.}

\author{B. Haas}
\affiliation{Univ. Grenoble Alpes, CEA, INAC, F38054 Grenoble, France.}

\author{P.-H. Jouneau}
\affiliation{Univ. Grenoble Alpes, CEA, INAC, F38054 Grenoble, France.}

\author{C. B\"{a}uerle}
\affiliation{CNRS, Inst NEEL, F-38042 Grenoble, France.}
\affiliation{Univ. Grenoble Alpes, Inst NEEL, F-38042 Grenoble, France.}

\author{L.P. L\'{e}vy}
\affiliation{CNRS, Inst NEEL, F-38042 Grenoble, France.}
\affiliation{Univ. Grenoble Alpes, Inst NEEL, F-38042 Grenoble, France.}

\author{E. Orignac}
\affiliation{Univ Lyon, ENS de Lyon, Univ Claude Bernard, CNRS, Laboratoire de Physique, F-69342 Lyon, France.}

\author{D. Carpentier}
\affiliation{Univ Lyon, ENS de Lyon, Univ Claude Bernard, CNRS, Laboratoire de Physique, F-69342 Lyon, France.}

\author{P. Ballet}
\affiliation{Univ. Grenoble Alpes, CEA, LETI, MINATEC campus, F38054 Grenoble, France.}

\author{T. Meunier}
\affiliation{CNRS, Inst NEEL, F-38042 Grenoble, France.}
\affiliation{Univ. Grenoble Alpes, Inst NEEL, F-38042 Grenoble, France.}

\date{\today}

\begin{abstract}

We demonstrate evidences of electronic transport via topological Dirac surface states in a thin film of strained HgTe. At high perpendicular magnetic fields, we show that the electron transport reaches the quantum Hall regime with vanishing resistance. Furthermore, quantum Hall transport spectroscopy reveals energy splittings of relativistic Landau levels specific to coupled Dirac surface states. This study provides new insights in the quantum Hall effect of topological insulator (TI) slabs, in the cross-over regime between two- and three-dimensional TIs, and in the relevance of thin TI films to explore novel circuit functionalities in spintronics and quantum nanoelectronics.

\end{abstract}


\maketitle 

Similar to the case of graphene, the charge carriers at the surface of topological insulators are expected to be massless Dirac fermions but with a real spin locked to the momentum \cite{Kane2007,Moore2007,Roy2009}. This has strong implications when the electrons experience a large perpendicular magnetic field $B$ and enter the quantum Hall regime. Indeed, each surface is then characterized by non-degenerate Landau levels (LL) and the associated Hall conductance is expressed as $\sigma_{xy}=(N+ \frac{1}{2}) \frac{e^2}{h}$, where $N$ is the LL index. In topological insulator slabs, however, two surface states of extension $w$ have to be considered and are separated by a thickness $t$. When the wave functions of the two surfaces do not overlap ($t \gg w$), they are only connected at the boundaries of the sample \cite{Lee2009} and the transport properties are obtained by summing the distinct contributions of each surface. The Hall conductance $\sigma_{xy}=(N_{top}+N_{bottom} +1)\frac{e^2}{h} = \nu \frac{e^2}{h}$  is then expected where $N_{top}$ and $N_{bottom}$ are the LL index for the top and bottom surfaces, respectively. In this regime, integer filling factors $\nu$ have been observed in relatively thick strained HgTe \cite{Brune2011,Kozlov2014,Kozlov2016} and Bi-based \cite{Sacepe2011,Xu2014} topological insulators. By decreasing $t$ down to $w$, the two surface states start to overlap giving rise to a non-negligible hybridization energy $\Delta$ \cite{Linder2009,Liu2010,Lu2010} and the transport then occurs through states delocalized between the two surfaces. Degenerate Dirac LLs are then expected to emerge at high $B$ with energies scaling as $\sqrt{NB}$. Moreover, additional dispersive couplings between the two surfaces are expected to lift the LL degeneracy with an energy splitting linear in $B$ \cite{Lu2010,Zhang2015}. While of orbital nature, this splitting reveals the microscopic coupling of spin and orbital degrees of freedom in the Dirac Landau levels. Therefore analysing the energy gaps of both odd and even filling factors and their $B$-dependences provides a powerful tool to reveal the Dirac surface states of a thin topological insulator slab. Decreasing further $t$ would result in the opening of a large gap in the surface states and the emergence of the quantum spin Hall phase \cite{Liu2010,Lu2010,Zhang2015,Konig2007}.

\begin{figure}\begin{center}
\includegraphics[scale=0.9]{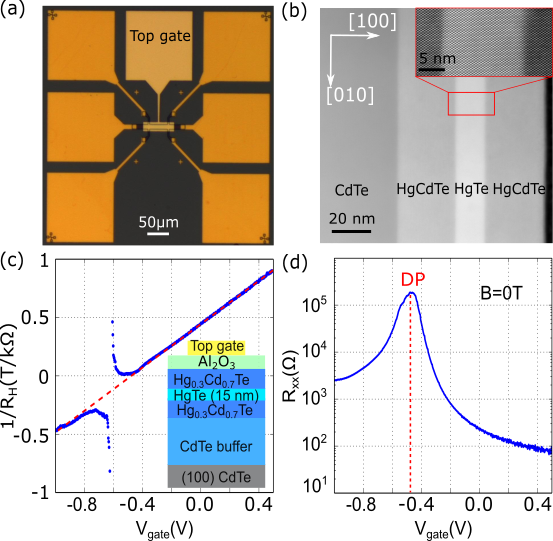} 
\caption{(a) Optical image of the Hall bar sample realized after nanofabrication of the structure. (b) Scanning Transmission Electron Microscopy (STEM) image in a High Angle Annular Dark Field (HAADF) mode of a 15 nm-thick HgTe layer embedded between two Hg$_{0.3}$Cd$_{0.7}$Te barriers. Inset: High resolution STEM HAADF image zooming on the HgTe layer. Low defect density and quality of the interfaces are evidenced.  (c) Evolution of the slope $R_H =\frac{dR_{xy}}{dB}$ at low perpendicular magnetic fields with the voltage $V_{gate}$ applied on the gate. In a one-carrier model, the density $n$ is equal to $\frac{1}{eR_H}$ where $e$ is the charge of an electron. $V_{gate}$ adjusts $n$ in the range of order of $10^{11}$ cm$^{-2}$. The red dashed line corresponds to a fit, which allows to extract the depleting factor of the gate $\alpha$ equal to $\sim 5.10^{11}$ cm$^{-2}$.V$^{-1}$. Inset: Schematics of the strained HgTe topological insulator structure used in the experiment. (d) Longitudinal resistance $R_{xx}$ as a function of $V_{gate}$ at zero magnetic field showing the Dirac point at $V_{gate} = V_{DP} \sim -0.5$V. } \label{fig1}
\end{center} \end{figure}

In this letter, the magneto-transport properties of strained HgTe thin films at high magnetic fields are investigated. We study films with a thickness of about 15 nm, characterized by two tunnel-coupled surfaces, where electron transport is solely mediated by surface charge carriers. The quality of the material allows reaching the quantum Hall regime with vanishing resistance for magnetic field larger than 1.5 T. At higher magnetic fields, non-degenerate Landau levels are observed. By analysing the temperature dependence of the magneto-conductance, a clear difference of the energy gaps corresponding to odd and even filling factors is noticed and is consistent with a Landau level energy spectrum characteristic of two coupled Dirac surfaces.
 
We investigate top-gated Hall bars (see Fig.\ref{fig1} (a)) fabricated from a 15 nm-thick HgTe layer surrounded by two 30 nm-thick Hg$_{0.3}$Cd$_{0.7}$Te barriers and grown on a (100) CdTe substrate. Particular attention to lower the defects present in the HgTe layer and to obtain sharp HgTe / Hg$_{0.3}$Cd$_{0.7}$Te interfaces was paid during the structure growth \cite{Ballet2014} (see Fig.\ref{fig1} (b)). Two different structures with similar thickness were grown and gave very similar results. The Hall bar is 40 $\mu$m long and 10 $\mu$m wide. A top gate covering the Hall bar enables to change the carriers from holes to electrons as illustrated by the density sign inversion in Fig.\ref{fig1} (c). 

In HgTe layers, the light hole band $\Gamma_{8,LH}$ band is lying 0.3 eV above the $\Gamma_{6}$. Such an inverted band structure at the $\Gamma$ point results in topological surface states, robust to the presence of the heavy hole band $\Gamma_{8,HH}$ \cite{Crauste2013}. At zero magnetic field and close to the charge neutrality point, the longitudinal resistance $R_{xx}$ presents a peak at gate voltage $V_{DP}$ (see Fig. \ref{fig1} (d)), whose amplitude depends on the size of the sample and can be as low as 1 k$\Omega$ for Hall bars of 1 micrometer \cite{suppmat}. We conclude that the structure has a metallic behavior as expected for electron transport through surface states.

To probe the nature of the surface state carriers, we analyse the Hall bar magneto-conductance at high perpendicular magnetic fields and at a temperature of 100 mK. Shubnikov-de Haas (SdH) oscillations on $R_{xx}$ and quantized plateaus on the Hall resistance $R_{xy}$ are observed (see Fig. \ref{fig2}(a) and (b)) and point at the emergence of LLs in the structure. Both $B$ and the top gate voltage $V_{gate}$ allow controlling the filling of individual LLs. Indeed, the carrier density $n$ is directly related to $B$ and $V_{gate}$ through  $n=\alpha V_{gate}= \nu \frac{eB}{h}$ where $\alpha$ is the depleting factor of the gate (see Fig. \ref{fig1} (c)). Derived from the extracted electron density at low $B$, the expected positions of the $R_{xx}$ minima are represented by the black dashed lines in Fig. \ref{fig2} (a). They are reproducing properly the minima of the fan diagram considering only odd filling factors on the hole-side ($V_{gate} \leq V_{DP}$) and both even and odd integers on the electron-side ($V_{gate} \geq V_{DP}$), in agreement with the corresponding quantized plateaus on the $R_{xy}$ mapping on Fig. \ref{fig2} (b).

\begin{figure}\begin{center}
\includegraphics[scale=1.0]{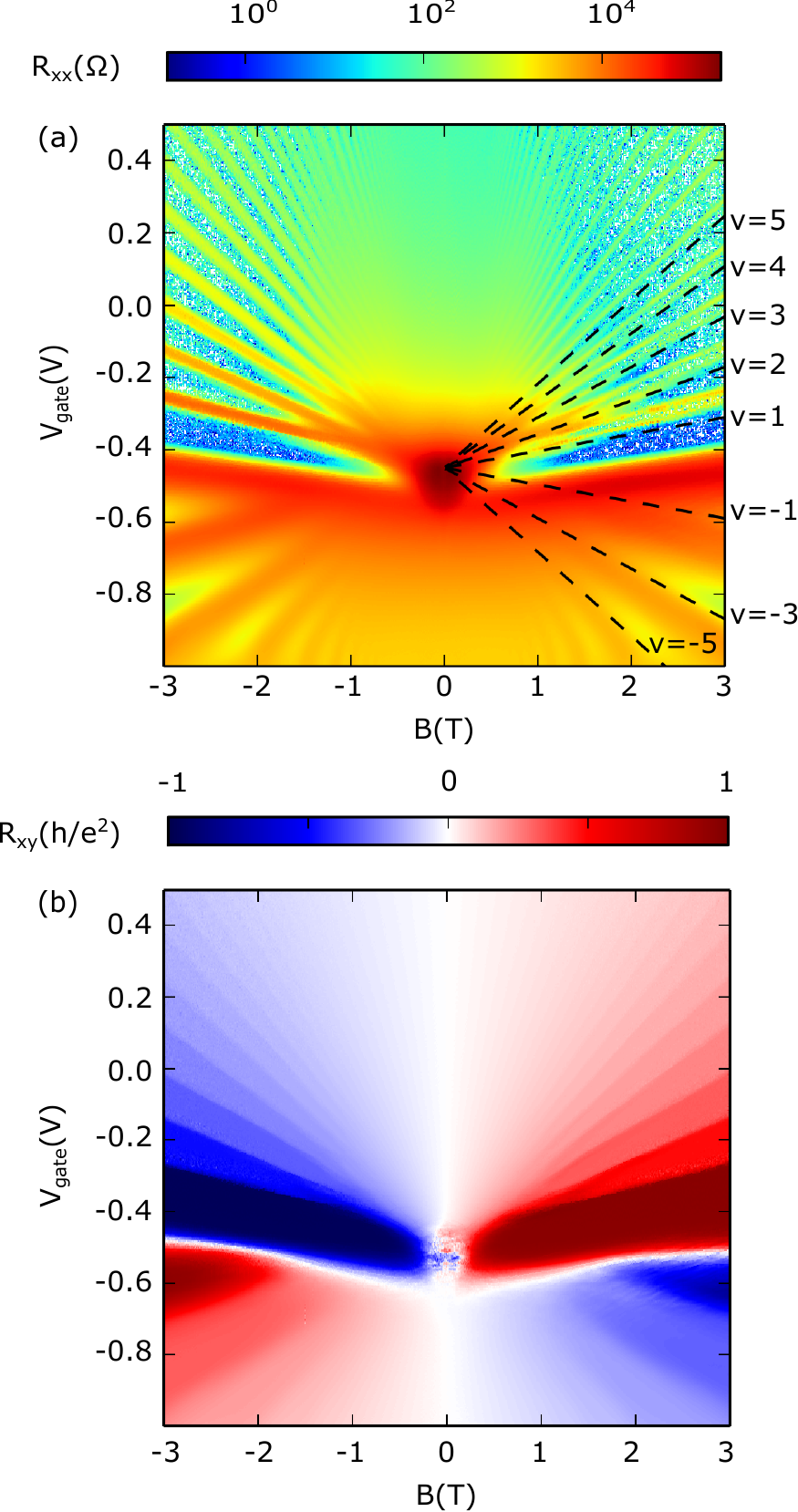} 
\caption{(a) Longitudinal resistance $R_{xx}$ as a function of $V_{gate}$ and the perpendicular magnetic field $B$ at 100 mK. The dashed lines represent the expected position of the $R_{xx}$ minima extracted from the density evolution with $V_{gate}$ (see Fig.\ref{fig1} (c)). (b) Hall resistance $R_{xy}$ (in units of $\frac{h^2}{e}$) as a function of $V_{gate}$ and $B$.} \label{fig2}
\end{center} \end{figure}

At $B = \pm$ 3 T, the quantum Hall regime is achieved on the electron-side with vanishing resistance and Hall conductance plateaus corresponding to integer filling factors (see Fig. \ref{fig3}). Such observations are strong signatures that there is no extra bulk contribution to the transport in the electron-regime in contrary to what is observed in thicker samples \cite{Brune2011,Kozlov2014,Kozlov2016}. On the hole-side, $\sigma_{xy}$  plateaus corresponding to odd filling factors are observed with $\sigma_{xx}$ no longer completely vanishing. Moreover, it is worth noticing that $\sigma_{xx}$ is characterized by broader peaks than on the electron-side.

The observed differences between holes and electrons are explained by the coupling of the surface states with the heavy hole $\Gamma_{8,HH}$ bulk band. From the band structure of strained HgTe \cite{Crauste2013}, the $\Gamma_{8,HH}$ band is expected to efficiently couple to the hole part of the surface states. This coupling opens up scattering channels resulting in the broadening of $\sigma_{xx}$ peaks. As a consequence, a larger magnetic field is needed on the hole-side to resolve the observed spin splitting on the electron-side. Whereas the $\nu$=2 plateau  is obtained for $|B|$ larger than $\sim$ 1.5 T  with a clear separation of the $R_{xx}$ maxima into two distinct branches (see Fig. \ref{fig2} (a)), additional measurements allow detecting the $\nu$ = -2 plateau on the hole-side only when $B$ equals 5.5 T \cite{suppmat}.
We can thus conclude that only one set of fan diagrams, associated to the same surface state on both electron- and hole-sides, is observed in the quantum Hall regime. Moreover, the degeneracy of the Landau levels is lifted for a 15 nm-thick HgTe topological film at high perpendicular magnetic fields.

\begin{figure}[!h]\begin{center}
\includegraphics[scale=1]{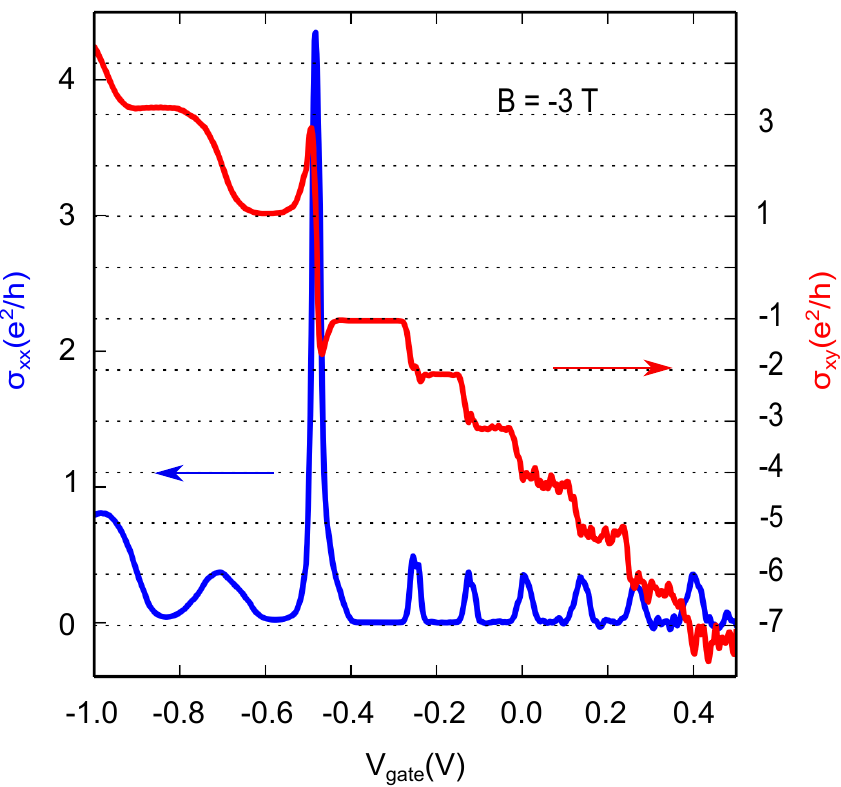} 
\caption{ The Hall $\sigma_{xy}= \frac{\rho_{xy}}{\rho_{xy}^2 + \rho_{xx}^2}$  and the longitudinal $\sigma_{xx}= \frac{\rho_{xx}}{\rho_{xy}^2 + \rho_{xx}^2}$  conductances as a function of $V_{gate}$ for $B$= -3 T with $\rho_{xy}=R_{xy}$ and $\rho_{xx}= R_{xx} \frac{W}{L}$, $L$ and $W$ respectively the length and the width of the Hall bar.} \label{fig3}
\end{center} \end{figure}

 To gain knowledge on the nature of the Landau levels, we analysed the temperature dependence of the magneto-conductance at high fields \cite{Jiang2007,Giesbers2007,Takase2015}. In such a procedure, the energy difference between LLs is estimated via thermal activation. More precisely, the temperature dependence of the longitudinal resistance $R_{xx}$ minima is fitted using the Arrhenius law (see Fig. \ref{fig4}(a) for $\nu=1$): $R_{xx}^{min}  \propto \exp{(\frac{-\Delta E}{2k_B T})}$ where $\Delta E$ is the activation energy gap, $k_B$ the Boltzmann constant and $T$ the temperature.  This analysis has only been performed on the electron-side since the influence of the $\Gamma_{8,HH}$ bulk band on the hole-side changes drastically with temperature \cite{suppmat}. We first focus on odd filling factors where two successive LLs have different orbitals (see Fig. \ref{fig4}(b) and (c)). We clearly observe non-regular energy separations between the successive LLs. Indeed, they decreases with $N$ and increases non-linearly with $B$ (mostly noticed for the $\nu$ = 1 gap). Both observations are in qualitative agreement with a Dirac-like LL energy spectrum scaling as $\sqrt{NB}$.

The resulting energy gaps for even filling factors do not show the same behavior (see Fig. \ref{fig4} (d)). Indeed, the activation energy gaps for all these LLs are similar and linear with $B$. The slope $2\beta$ has been estimated to be about $2.07 \pm 0.65$ meV/T from the red solid line fit of Fig. \ref{fig4} (d), which is similar to the Landau level splitting observed in HgTe non-topological quantum wells \cite{Gui2004,Buttner2011}.

To quantitatively model the energy gaps for odd and even filling factors, we consider an effective low-energy model of a thin 3D topological insulator: two Dirac surface states of opposite chirality $\hbar v_f \tau_z (\sigma_x .k_y - \sigma_y . k_x) $   coupled by a $k$-dependent hopping amplitude of $(\frac{\Delta}{2}- \frac{k^2}{2M})\tau_x$ where $v_f$ is the surface state band velocity equal to $5.10^5$ m.s$^{-1}$ in our structures \cite{Crauste2013}. This coupling is purely of orbital nature, it does not depend on spins and originates from the overlap of the two surface states that opens a constant gap $\Delta$  and introduces a quadratic $\frac{k^2}{2M}$ hopping amplitude \cite{Lu2010}. In these conditions, we obtain the following LL energy spectrum with $\beta=\frac{e}{2M}$ explained in further details in ref \cite{suppmat}.

\begin{equation}
E_{N, \pm}= \pm \sqrt{2Ne\hbar v_f^2 B+(\frac{\Delta}{2}-N\beta B)^2 } \pm \beta B    \textrm{,}   N \geq 1
\label{eq_1}
\end{equation}

\begin{equation}
E_{0, \pm}= \pm |\beta B -\frac{\Delta}{2}|
\label{eq_2}
\end{equation}

\begin{figure}\begin{center}
\includegraphics[scale=1]{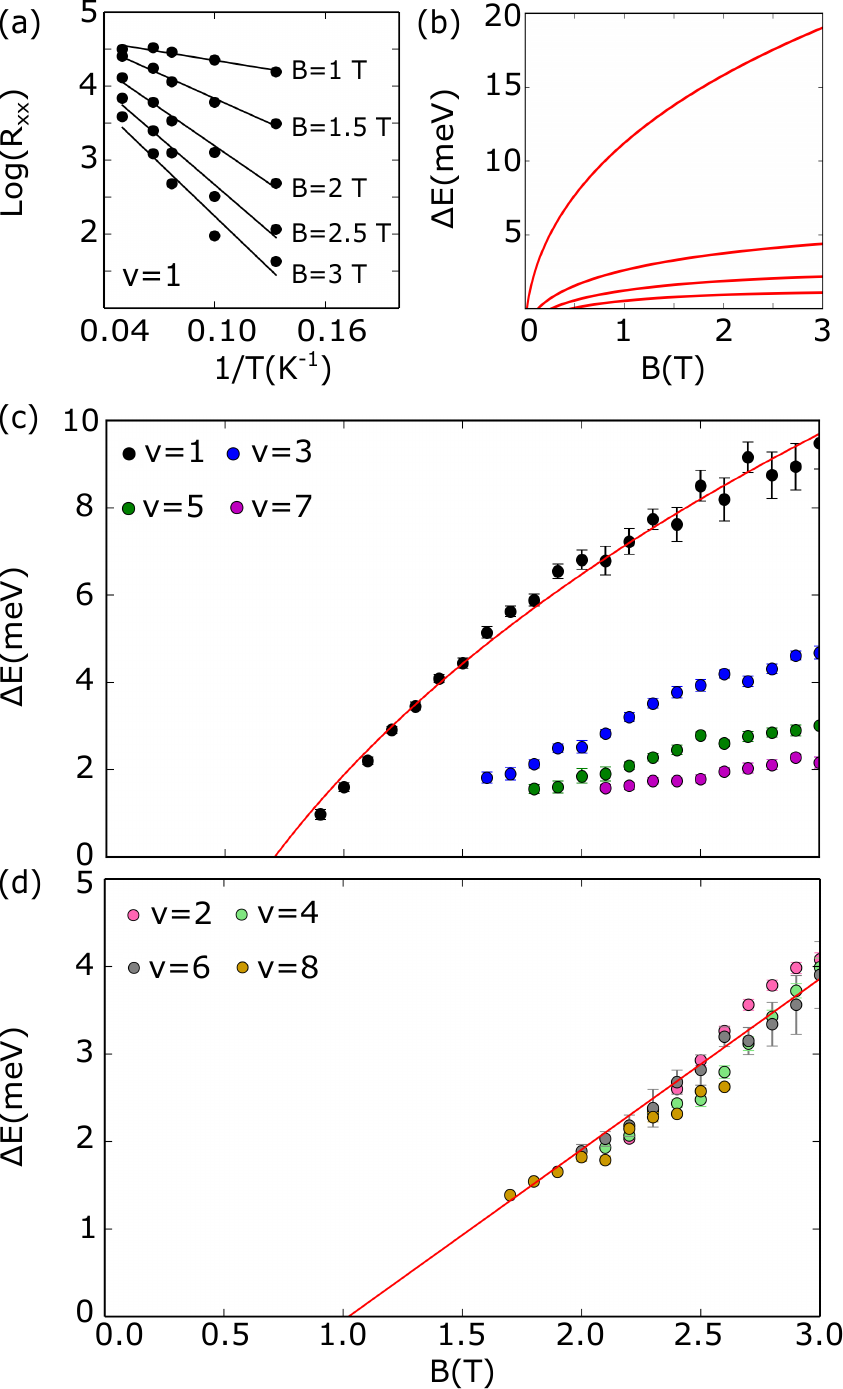} 
\caption{Minimum of $R_{xx}$ as a function of temperature $T$ for the $\nu$=1 filling factor (a). This analysis allows to extract the activation energy gap $\Delta E$ for all positive filling factors at different $B$. $\Delta E$ evolution with $B$ 
for odd filling factors $\nu=1, 3, 5$ and $7$: (b) from the model of eq. \ref{eq_1} and \ref{eq_2}: $\Delta E_{\nu=2N+1}=E_{N+1,-}- E_{N,+}$, (c) from the experimental data. (d) $\Delta E$ as a function of $B$ for even filling factors. 
 In (c) and (d), the red solid lines represent the result of the model with a constant offset $\Gamma$ (see text and ref. \cite{suppmat}). 
 In (c), $\Delta E_{\nu=2N+1}=E_{N+1,-}- E_{N,+}+\Gamma$ with $\Gamma=11$ meV. 
 In (d), $\Delta E_{\nu=2N} = E_{N,+}-E_{N,-}+ \Gamma$ with $\Gamma=3$ meV. } \label{fig4}
\end{center} \end{figure}

While of orbital nature, the term $\beta B$ acts as an effective Zeeman energy and is fixed by the analysis of the even filling factor energy gaps performed in the previous paragraph. This spectrum evidences a splitting of each orbital LL due to the presence of the quadratic term. It is worth noting that for a vanishing overlap between the two surfaces both the gap and the quadratic term go to zero and one should observe only odd filling factors \cite{Lu2010}. The results of this model for the estimated $v_f$ are presented in Fig. \ref{fig4}(b), (c) and (d) through the solid red lines. To obtain a good agreement between the data and the model for all energy gaps, we need to take into account the width of the Landau level $\Gamma$ as a constant offset in energy, estimated to 3 meV from the SDH oscillations on the electron-side \cite{suppmat}. Nevertheless, for $\nu$ = 1, the model requires a three times larger $\Gamma$ (solid red line in Fig. \ref{fig4} (c)) that can be explained by the proximity in energy of the $\Gamma_{8,HH}$ band \cite{suppmat}. These analysis point at the Dirac nature of the surface-state carriers of strained HgTe thin films.\\

In conclusion, we have shown that the quantum Hall regime with vanishing longitudinal resistance is achieved in 15 nm-thick strained HgTe layers. Dirac carriers are revealed via quantum Hall spectroscopy and propagate solely through states delocalized between the top and bottom topological surfaces of the HgTe slab. The overlap between the surface states results in the mixing between the two Dirac species thus in the splitting of the Landau levels. Dirac-surface-restricted transport opens the route towards the realization of topological circuits based on strained HgTe thin films, with potential applications in quantum nanoelectronics\cite{Jiang2011} and spintronics\cite{Rojas2016}.\\

We acknowledge  N. Mollard for TEM sample preparation and the technical support from the technological centres of the Institut N\'eel, CEA-LETI and PTA. T.M. acknowledges financial support from Fondation Nanosciences and ANR Semitopo.

\bibliographystyle{apsrev4-1}
\bibliography{bib}

\begin{thebibliography}{24}%
\makeatletter
\providecommand \@ifxundefined [1]{%
 \@ifx{#1\undefined}
}%
\providecommand \@ifnum [1]{%
 \ifnum #1\expandafter \@firstoftwo
 \else \expandafter \@secondoftwo
 \fi
}%
\providecommand \@ifx [1]{%
 \ifx #1\expandafter \@firstoftwo
 \else \expandafter \@secondoftwo
 \fi
}%
\providecommand \natexlab [1]{#1}%
\providecommand \enquote  [1]{``#1''}%
\providecommand \bibnamefont  [1]{#1}%
\providecommand \bibfnamefont [1]{#1}%
\providecommand \citenamefont [1]{#1}%
\providecommand \href@noop [0]{\@secondoftwo}%
\providecommand \href [0]{\begingroup \@sanitize@url \@href}%
\providecommand \@href[1]{\@@startlink{#1}\@@href}%
\providecommand \@@href[1]{\endgroup#1\@@endlink}%
\providecommand \@sanitize@url [0]{\catcode `\\12\catcode `\$12\catcode
  `\&12\catcode `\#12\catcode `\^12\catcode `\_12\catcode `\%12\relax}%
\providecommand \@@startlink[1]{}%
\providecommand \@@endlink[0]{}%
\providecommand \url  [0]{\begingroup\@sanitize@url \@url }%
\providecommand \@url [1]{\endgroup\@href {#1}{\urlprefix }}%
\providecommand \urlprefix  [0]{URL }%
\providecommand \Eprint [0]{\href }%
\providecommand \doibase [0]{http://dx.doi.org/}%
\providecommand \selectlanguage [0]{\@gobble}%
\providecommand \bibinfo  [0]{\@secondoftwo}%
\providecommand \bibfield  [0]{\@secondoftwo}%
\providecommand \translation [1]{[#1]}%
\providecommand \BibitemOpen [0]{}%
\providecommand \bibitemStop [0]{}%
\providecommand \bibitemNoStop [0]{.\EOS\space}%
\providecommand \EOS [0]{\spacefactor3000\relax}%
\providecommand \BibitemShut  [1]{\csname bibitem#1\endcsname}%
\let\auto@bib@innerbib\@empty
\bibitem [{\citenamefont {Kane}\ and\ \citenamefont {Mele}(2007)}]{Kane2007}%
  \BibitemOpen
  \bibfield  {author} {\bibinfo {author} {\bibfnamefont {C.}~\bibnamefont
  {Kane}}\ and\ \bibinfo {author} {\bibfnamefont {E.}~\bibnamefont {Mele}},\
  }\href@noop {} {\bibfield  {journal} {\bibinfo  {journal} {Phys. Rev. Lett.}\
  }\textbf {\bibinfo {volume} {98}},\ \bibinfo {pages} {106803} (\bibinfo
  {year} {2007})}\BibitemShut {NoStop}%
\bibitem [{\citenamefont {Moore}\ and\ \citenamefont
  {Balents}(2007)}]{Moore2007}%
  \BibitemOpen
  \bibfield  {author} {\bibinfo {author} {\bibfnamefont {J.}~\bibnamefont
  {Moore}}\ and\ \bibinfo {author} {\bibfnamefont {L.}~\bibnamefont
  {Balents}},\ }\href@noop {} {\bibfield  {journal} {\bibinfo  {journal} {Phys.
  Rev. B}\ }\textbf {\bibinfo {volume} {75}},\ \bibinfo {pages} {121306(R)}
  (\bibinfo {year} {2007})}\BibitemShut {NoStop}%
\bibitem [{\citenamefont {Roy}(2009)}]{Roy2009}%
  \BibitemOpen
  \bibfield  {author} {\bibinfo {author} {\bibfnamefont {R.}~\bibnamefont
  {Roy}},\ }\href@noop {} {\bibfield  {journal} {\bibinfo  {journal} {Phys.
  Rev. B}\ }\textbf {\bibinfo {volume} {79}},\ \bibinfo {pages} {195322}
  (\bibinfo {year} {2009})}\BibitemShut {NoStop}%
\bibitem [{\citenamefont {Lee}(2009)}]{Lee2009}%
  \BibitemOpen
  \bibfield  {author} {\bibinfo {author} {\bibfnamefont {D.}~\bibnamefont
  {Lee}},\ }\href@noop {} {\bibfield  {journal} {\bibinfo  {journal} {Phys.
  Rev. Lett.}\ }\textbf {\bibinfo {volume} {103}},\ \bibinfo {pages} {196804}
  (\bibinfo {year} {2009})}\BibitemShut {NoStop}%
\bibitem [{\citenamefont {Br\"{u}ne}\ \emph {et~al.}(2011)\citenamefont
  {Br\"{u}ne}, \citenamefont {Liu}, \citenamefont {Novik}, \citenamefont
  {Hankiewicz}, \citenamefont {Buhmann}, \citenamefont {Chen}, \citenamefont
  {Qi}, \citenamefont {Shen}, \citenamefont {Zhang},\ and\ \citenamefont
  {Molenkamp}}]{Brune2011}%
  \BibitemOpen
  \bibfield  {author} {\bibinfo {author} {\bibfnamefont {C.}~\bibnamefont
  {Br\"{u}ne}}, \bibinfo {author} {\bibfnamefont {C.~X.}\ \bibnamefont {Liu}},
  \bibinfo {author} {\bibfnamefont {E.~G.}\ \bibnamefont {Novik}}, \bibinfo
  {author} {\bibfnamefont {E.~M.}\ \bibnamefont {Hankiewicz}}, \bibinfo
  {author} {\bibfnamefont {H.}~\bibnamefont {Buhmann}}, \bibinfo {author}
  {\bibfnamefont {Y.~L.}\ \bibnamefont {Chen}}, \bibinfo {author}
  {\bibfnamefont {X.~L.}\ \bibnamefont {Qi}}, \bibinfo {author} {\bibfnamefont
  {Z.~X.}\ \bibnamefont {Shen}}, \bibinfo {author} {\bibfnamefont {S.~C.}\
  \bibnamefont {Zhang}}, \ and\ \bibinfo {author} {\bibfnamefont {L.~W.}\
  \bibnamefont {Molenkamp}},\ }\href@noop {} {\bibfield  {journal} {\bibinfo
  {journal} {Phys. Rev. Lett.}\ }\textbf {\bibinfo {volume} {106}},\ \bibinfo
  {pages} {126803} (\bibinfo {year} {2011})}\BibitemShut {NoStop}%
\bibitem [{\citenamefont {Kozlov}\ \emph {et~al.}(2014)\citenamefont {Kozlov},
  \citenamefont {Kvon}, \citenamefont {Olshanetsky}, \citenamefont {Mikhailov},
  \citenamefont {Dvoretsky},\ and\ \citenamefont {Weiss}}]{Kozlov2014}%
  \BibitemOpen
  \bibfield  {author} {\bibinfo {author} {\bibfnamefont {D.~A.}\ \bibnamefont
  {Kozlov}}, \bibinfo {author} {\bibfnamefont {Z.~D.}\ \bibnamefont {Kvon}},
  \bibinfo {author} {\bibfnamefont {E.~B.}\ \bibnamefont {Olshanetsky}},
  \bibinfo {author} {\bibfnamefont {N.~N.}\ \bibnamefont {Mikhailov}}, \bibinfo
  {author} {\bibfnamefont {S.~A.}\ \bibnamefont {Dvoretsky}}, \ and\ \bibinfo
  {author} {\bibfnamefont {D.}~\bibnamefont {Weiss}},\ }\href@noop {}
  {\bibfield  {journal} {\bibinfo  {journal} {Phys. Rev. Lett.}\ }\textbf
  {\bibinfo {volume} {112}},\ \bibinfo {pages} {196801} (\bibinfo {year}
  {2014})}\BibitemShut {NoStop}%
\bibitem [{\citenamefont {Kozlov}\ \emph {et~al.}(2016)\citenamefont {Kozlov},
  \citenamefont {Bauer}, \citenamefont {Ziegler}, \citenamefont {Fischer},
  \citenamefont {Savchenko}, \citenamefont {Kvon}, \citenamefont {Mikhailov},
  \citenamefont {Dvoretsky},\ and\ \citenamefont {Weiss}}]{Kozlov2016}%
  \BibitemOpen
  \bibfield  {author} {\bibinfo {author} {\bibfnamefont {D.~A.}\ \bibnamefont
  {Kozlov}}, \bibinfo {author} {\bibfnamefont {D.}~\bibnamefont {Bauer}},
  \bibinfo {author} {\bibfnamefont {J.}~\bibnamefont {Ziegler}}, \bibinfo
  {author} {\bibfnamefont {R.}~\bibnamefont {Fischer}}, \bibinfo {author}
  {\bibfnamefont {M.~L.}\ \bibnamefont {Savchenko}}, \bibinfo {author}
  {\bibfnamefont {Z.~D.}\ \bibnamefont {Kvon}}, \bibinfo {author}
  {\bibfnamefont {N.~N.}\ \bibnamefont {Mikhailov}}, \bibinfo {author}
  {\bibfnamefont {S.~A.}\ \bibnamefont {Dvoretsky}}, \ and\ \bibinfo {author}
  {\bibfnamefont {D.}~\bibnamefont {Weiss}},\ }\href@noop {} {\bibfield
  {journal} {\bibinfo  {journal} {Phys. Rev. Lett.}\ }\textbf {\bibinfo
  {volume} {116}},\ \bibinfo {pages} {166802} (\bibinfo {year}
  {2016})}\BibitemShut {NoStop}%
\bibitem [{\citenamefont {Sac\'{e}p\'{e}}\ \emph {et~al.}(2011)\citenamefont
  {Sac\'{e}p\'{e}}, \citenamefont {Oostinga}, \citenamefont {Li}, \citenamefont
  {Ubaldini}, \citenamefont {Couto}, \citenamefont {Giannini},\ and\
  \citenamefont {Morpurgo}}]{Sacepe2011}%
  \BibitemOpen
  \bibfield  {author} {\bibinfo {author} {\bibfnamefont {B.}~\bibnamefont
  {Sac\'{e}p\'{e}}}, \bibinfo {author} {\bibfnamefont {J.}~\bibnamefont
  {Oostinga}}, \bibinfo {author} {\bibfnamefont {J.}~\bibnamefont {Li}},
  \bibinfo {author} {\bibfnamefont {A.}~\bibnamefont {Ubaldini}}, \bibinfo
  {author} {\bibfnamefont {N.}~\bibnamefont {Couto}}, \bibinfo {author}
  {\bibfnamefont {E.}~\bibnamefont {Giannini}}, \ and\ \bibinfo {author}
  {\bibfnamefont {A.}~\bibnamefont {Morpurgo}},\ }\href@noop {} {\bibfield
  {journal} {\bibinfo  {journal} {Nature Communications}\ }\textbf {\bibinfo
  {volume} {2}},\ \bibinfo {pages} {575} (\bibinfo {year} {2011})}\BibitemShut
  {NoStop}%
\bibitem [{\citenamefont {Xu}\ \emph {et~al.}(2014)\citenamefont {Xu},
  \citenamefont {Miotkowski}, \citenamefont {Liu}, \citenamefont {Tian},
  \citenamefont {Nam}, \citenamefont {Alidoust}, \citenamefont {Hu},
  \citenamefont {Shih}, \citenamefont {Hasan},\ and\ \citenamefont
  {Chen}}]{Xu2014}%
  \BibitemOpen
  \bibfield  {author} {\bibinfo {author} {\bibfnamefont {Y.}~\bibnamefont
  {Xu}}, \bibinfo {author} {\bibfnamefont {I.}~\bibnamefont {Miotkowski}},
  \bibinfo {author} {\bibfnamefont {C.}~\bibnamefont {Liu}}, \bibinfo {author}
  {\bibfnamefont {J.}~\bibnamefont {Tian}}, \bibinfo {author} {\bibfnamefont
  {H.}~\bibnamefont {Nam}}, \bibinfo {author} {\bibfnamefont {N.}~\bibnamefont
  {Alidoust}}, \bibinfo {author} {\bibfnamefont {J.}~\bibnamefont {Hu}},
  \bibinfo {author} {\bibfnamefont {C.-K.}\ \bibnamefont {Shih}}, \bibinfo
  {author} {\bibfnamefont {M.}~\bibnamefont {Hasan}}, \ and\ \bibinfo {author}
  {\bibfnamefont {Y.}~\bibnamefont {Chen}},\ }\href@noop {} {\bibfield
  {journal} {\bibinfo  {journal} {Nature Physics}\ }\textbf {\bibinfo {volume}
  {10}},\ \bibinfo {pages} {956} (\bibinfo {year} {2014})}\BibitemShut
  {NoStop}%
\bibitem [{\citenamefont {Linder}\ \emph {et~al.}(2009)\citenamefont {Linder},
  \citenamefont {Yokoyama},\ and\ \citenamefont {Sudbo}}]{Linder2009}%
  \BibitemOpen
  \bibfield  {author} {\bibinfo {author} {\bibfnamefont {J.}~\bibnamefont
  {Linder}}, \bibinfo {author} {\bibfnamefont {T.}~\bibnamefont {Yokoyama}}, \
  and\ \bibinfo {author} {\bibfnamefont {A.}~\bibnamefont {Sudbo}},\
  }\href@noop {} {\bibfield  {journal} {\bibinfo  {journal} {Phys. Rev. B}\
  }\textbf {\bibinfo {volume} {80}},\ \bibinfo {pages} {205401} (\bibinfo
  {year} {2009})}\BibitemShut {NoStop}%
\bibitem [{\citenamefont {Liu}\ \emph {et~al.}(2010)\citenamefont {Liu},
  \citenamefont {Zhang}, \citenamefont {Yan}, \citenamefont {Qi}, \citenamefont
  {Frauenheim}, \citenamefont {Dai}, \citenamefont {Fang},\ and\ \citenamefont
  {Zhang}}]{Liu2010}%
  \BibitemOpen
  \bibfield  {author} {\bibinfo {author} {\bibfnamefont {C.-X.}\ \bibnamefont
  {Liu}}, \bibinfo {author} {\bibfnamefont {H.}~\bibnamefont {Zhang}}, \bibinfo
  {author} {\bibfnamefont {B.}~\bibnamefont {Yan}}, \bibinfo {author}
  {\bibfnamefont {X.-L.}\ \bibnamefont {Qi}}, \bibinfo {author} {\bibfnamefont
  {T.}~\bibnamefont {Frauenheim}}, \bibinfo {author} {\bibfnamefont
  {X.}~\bibnamefont {Dai}}, \bibinfo {author} {\bibfnamefont {Z.}~\bibnamefont
  {Fang}}, \ and\ \bibinfo {author} {\bibfnamefont {S.-C.}\ \bibnamefont
  {Zhang}},\ }\href@noop {} {\bibfield  {journal} {\bibinfo  {journal} {Phys.
  Rev. B}\ }\textbf {\bibinfo {volume} {81}},\ \bibinfo {pages} {041307}
  (\bibinfo {year} {2010})}\BibitemShut {NoStop}%
\bibitem [{\citenamefont {Lu}\ \emph {et~al.}(2010)\citenamefont {Lu},
  \citenamefont {Shan}, \citenamefont {Yao}, \citenamefont {Niu},\ and\
  \citenamefont {Shen}}]{Lu2010}%
  \BibitemOpen
  \bibfield  {author} {\bibinfo {author} {\bibfnamefont {H.-Z.}\ \bibnamefont
  {Lu}}, \bibinfo {author} {\bibfnamefont {W.-Y.}\ \bibnamefont {Shan}},
  \bibinfo {author} {\bibfnamefont {W.}~\bibnamefont {Yao}}, \bibinfo {author}
  {\bibfnamefont {Q.}~\bibnamefont {Niu}}, \ and\ \bibinfo {author}
  {\bibfnamefont {S.-Q.}\ \bibnamefont {Shen}},\ }\href@noop {} {\bibfield
  {journal} {\bibinfo  {journal} {Phys. Rev. B}\ }\textbf {\bibinfo {volume}
  {81}},\ \bibinfo {pages} {115407} (\bibinfo {year} {2010})}\BibitemShut
  {NoStop}%
\bibitem [{\citenamefont {Zhang}\ \emph {et~al.}(2015)\citenamefont {Zhang},
  \citenamefont {Lu},\ and\ \citenamefont {Shen}}]{Zhang2015}%
  \BibitemOpen
  \bibfield  {author} {\bibinfo {author} {\bibfnamefont {S.-B.}\ \bibnamefont
  {Zhang}}, \bibinfo {author} {\bibfnamefont {H.-Z.}\ \bibnamefont {Lu}}, \
  and\ \bibinfo {author} {\bibfnamefont {S.-Q.}\ \bibnamefont {Shen}},\
  }\href@noop {} {\bibfield  {journal} {\bibinfo  {journal} {Scientific
  Reports}\ }\textbf {\bibinfo {volume} {5}},\ \bibinfo {pages} {13277}
  (\bibinfo {year} {2015})}\BibitemShut {NoStop}%
\bibitem [{\citenamefont {K\"{o}nig}\ \emph {et~al.}(2007)\citenamefont
  {K\"{o}nig}, \citenamefont {Wiedmann}, \citenamefont {Br\"{u}ne},
  \citenamefont {Roth}, \citenamefont {Buhmann}, \citenamefont {Molenkamp},
  \citenamefont {Qi},\ and\ \citenamefont {Zhang}}]{Konig2007}%
  \BibitemOpen
  \bibfield  {author} {\bibinfo {author} {\bibfnamefont {M.}~\bibnamefont
  {K\"{o}nig}}, \bibinfo {author} {\bibfnamefont {S.}~\bibnamefont {Wiedmann}},
  \bibinfo {author} {\bibfnamefont {C.}~\bibnamefont {Br\"{u}ne}}, \bibinfo
  {author} {\bibfnamefont {A.}~\bibnamefont {Roth}}, \bibinfo {author}
  {\bibfnamefont {H.}~\bibnamefont {Buhmann}}, \bibinfo {author} {\bibfnamefont
  {L.}~\bibnamefont {Molenkamp}}, \bibinfo {author} {\bibfnamefont {X.-L.}\
  \bibnamefont {Qi}}, \ and\ \bibinfo {author} {\bibfnamefont {S.-C.}\
  \bibnamefont {Zhang}},\ }\href@noop {} {\bibfield  {journal} {\bibinfo
  {journal} {Science}\ }\textbf {\bibinfo {volume} {318}},\ \bibinfo {pages}
  {766} (\bibinfo {year} {2007})}\BibitemShut {NoStop}%
\bibitem [{\citenamefont {Ballet}\ \emph {et~al.}(2014)\citenamefont {Ballet},
  \citenamefont {Thomas}, \citenamefont {Baudry}, \citenamefont {Bouvier},
  \citenamefont {Crauste}, \citenamefont {Meunier}, \citenamefont {Badano},
  \citenamefont {Veillerot}, \citenamefont {Barnes}, \citenamefont {Jouneau},\
  and\ \citenamefont {L\'{e}vy}}]{Ballet2014}%
  \BibitemOpen
  \bibfield  {author} {\bibinfo {author} {\bibfnamefont {P.}~\bibnamefont
  {Ballet}}, \bibinfo {author} {\bibfnamefont {C.}~\bibnamefont {Thomas}},
  \bibinfo {author} {\bibfnamefont {X.}~\bibnamefont {Baudry}}, \bibinfo
  {author} {\bibfnamefont {C.}~\bibnamefont {Bouvier}}, \bibinfo {author}
  {\bibfnamefont {O.}~\bibnamefont {Crauste}}, \bibinfo {author} {\bibfnamefont
  {T.}~\bibnamefont {Meunier}}, \bibinfo {author} {\bibfnamefont
  {G.}~\bibnamefont {Badano}}, \bibinfo {author} {\bibfnamefont
  {M.}~\bibnamefont {Veillerot}}, \bibinfo {author} {\bibfnamefont {J.~P.}\
  \bibnamefont {Barnes}}, \bibinfo {author} {\bibfnamefont {P.}~\bibnamefont
  {Jouneau}}, \ and\ \bibinfo {author} {\bibfnamefont {L.}~\bibnamefont
  {L\'{e}vy}},\ }\href@noop {} {\bibfield  {journal} {\bibinfo  {journal} {J.
  Electron. Mater.}\ }\textbf {\bibinfo {volume} {43}},\ \bibinfo {pages}
  {2955} (\bibinfo {year} {2014})}\BibitemShut {NoStop}%
\bibitem [{\citenamefont {Crauste}\ \emph {et~al.}()\citenamefont {Crauste},
  \citenamefont {Ohtsubo}, \citenamefont {Ballet}, \citenamefont {Delplace},
  \citenamefont {Carpentier}, \citenamefont {Bouvier}, \citenamefont {Meunier},
  \citenamefont {Taleb-Ibrahimi},\ and\ \citenamefont
  {L\'{e}vy}}]{Crauste2013}%
  \BibitemOpen
  \bibfield  {author} {\bibinfo {author} {\bibfnamefont {O.}~\bibnamefont
  {Crauste}}, \bibinfo {author} {\bibfnamefont {Y.}~\bibnamefont {Ohtsubo}},
  \bibinfo {author} {\bibfnamefont {P.}~\bibnamefont {Ballet}}, \bibinfo
  {author} {\bibfnamefont {P.}~\bibnamefont {Delplace}}, \bibinfo {author}
  {\bibfnamefont {D.}~\bibnamefont {Carpentier}}, \bibinfo {author}
  {\bibfnamefont {C.}~\bibnamefont {Bouvier}}, \bibinfo {author} {\bibfnamefont
  {T.}~\bibnamefont {Meunier}}, \bibinfo {author} {\bibfnamefont
  {A.}~\bibnamefont {Taleb-Ibrahimi}}, \ and\ \bibinfo {author} {\bibfnamefont
  {L.~P.}\ \bibnamefont {L\'{e}vy}},\ }\href@noop {} {\bibinfo  {journal}
  {arXiv:1307.2008}\ }\BibitemShut {NoStop}%
\bibitem [{\citenamefont {see~supplementary materials}()}]{suppmat}%
  \BibitemOpen
\bibfield  {journal} {  }\bibfield  {author} {\bibinfo {author} {\bibnamefont
  {see~supplementary materials}},\ }\href@noop {} {\ }\BibitemShut {NoStop}%
\bibitem [{\citenamefont {Jiang}\ \emph {et~al.}(2007)\citenamefont {Jiang},
  \citenamefont {Zhang}, \citenamefont {Stormer},\ and\ \citenamefont
  {Kim}}]{Jiang2007}%
  \BibitemOpen
  \bibfield  {author} {\bibinfo {author} {\bibfnamefont {Z.}~\bibnamefont
  {Jiang}}, \bibinfo {author} {\bibfnamefont {Y.}~\bibnamefont {Zhang}},
  \bibinfo {author} {\bibfnamefont {H.~L.}\ \bibnamefont {Stormer}}, \ and\
  \bibinfo {author} {\bibfnamefont {P.}~\bibnamefont {Kim}},\ }\href@noop {}
  {\bibfield  {journal} {\bibinfo  {journal} {Phys. Rev. Lett.}\ }\textbf
  {\bibinfo {volume} {99}},\ \bibinfo {pages} {106802} (\bibinfo {year}
  {2007})}\BibitemShut {NoStop}%
\bibitem [{\citenamefont {Giesbers}\ \emph {et~al.}(2007)\citenamefont
  {Giesbers}, \citenamefont {Zeitler}, \citenamefont {Katsnelson},
  \citenamefont {Ponomarenko}, \citenamefont {Mohiuddin},\ and\ \citenamefont
  {Maan}}]{Giesbers2007}%
  \BibitemOpen
  \bibfield  {author} {\bibinfo {author} {\bibfnamefont {A.~J.~M.}\
  \bibnamefont {Giesbers}}, \bibinfo {author} {\bibfnamefont {U.}~\bibnamefont
  {Zeitler}}, \bibinfo {author} {\bibfnamefont {M.~I.}\ \bibnamefont
  {Katsnelson}}, \bibinfo {author} {\bibfnamefont {L.~A.}\ \bibnamefont
  {Ponomarenko}}, \bibinfo {author} {\bibfnamefont {T.~M.}\ \bibnamefont
  {Mohiuddin}}, \ and\ \bibinfo {author} {\bibfnamefont {J.~C.}\ \bibnamefont
  {Maan}},\ }\href@noop {} {\bibfield  {journal} {\bibinfo  {journal} {Phys.
  Rev. Lett.}\ }\textbf {\bibinfo {volume} {99}},\ \bibinfo {pages} {206803}
  (\bibinfo {year} {2007})}\BibitemShut {NoStop}%
\bibitem [{\citenamefont {Takase}\ \emph {et~al.}(2015)\citenamefont {Takase},
  \citenamefont {Hibino},\ and\ \citenamefont {Muraki}}]{Takase2015}%
  \BibitemOpen
  \bibfield  {author} {\bibinfo {author} {\bibfnamefont {K.}~\bibnamefont
  {Takase}}, \bibinfo {author} {\bibfnamefont {H.}~\bibnamefont {Hibino}}, \
  and\ \bibinfo {author} {\bibfnamefont {K.}~\bibnamefont {Muraki}},\
  }\href@noop {} {\bibfield  {journal} {\bibinfo  {journal} {Phys. Rev. B.}\
  }\textbf {\bibinfo {volume} {92}},\ \bibinfo {pages} {125407} (\bibinfo
  {year} {2015})}\BibitemShut {NoStop}%
\bibitem [{\citenamefont {Gui}\ \emph {et~al.}(2004)\citenamefont {Gui},
  \citenamefont {Becker}, \citenamefont {Dai}, \citenamefont {Liu},
  \citenamefont {Qiu}, \citenamefont {Novik}, \citenamefont {Sch\"afer},
  \citenamefont {Shu}, \citenamefont {Chu}, \citenamefont {Buhmann},\ and\
  \citenamefont {Molenkamp}}]{Gui2004}%
  \BibitemOpen
  \bibfield  {author} {\bibinfo {author} {\bibfnamefont {Y.~S.}\ \bibnamefont
  {Gui}}, \bibinfo {author} {\bibfnamefont {C.~R.}\ \bibnamefont {Becker}},
  \bibinfo {author} {\bibfnamefont {N.}~\bibnamefont {Dai}}, \bibinfo {author}
  {\bibfnamefont {J.}~\bibnamefont {Liu}}, \bibinfo {author} {\bibfnamefont
  {Z.~J.}\ \bibnamefont {Qiu}}, \bibinfo {author} {\bibfnamefont {E.~G.}\
  \bibnamefont {Novik}}, \bibinfo {author} {\bibfnamefont {M.}~\bibnamefont
  {Sch\"afer}}, \bibinfo {author} {\bibfnamefont {X.~Z.}\ \bibnamefont {Shu}},
  \bibinfo {author} {\bibfnamefont {J.~H.}\ \bibnamefont {Chu}}, \bibinfo
  {author} {\bibfnamefont {H.}~\bibnamefont {Buhmann}}, \ and\ \bibinfo
  {author} {\bibfnamefont {L.~W.}\ \bibnamefont {Molenkamp}},\ }\href@noop {}
  {\bibfield  {journal} {\bibinfo  {journal} {Phys. Rev. B}\ }\textbf {\bibinfo
  {volume} {70}},\ \bibinfo {pages} {115328} (\bibinfo {year}
  {2004})}\BibitemShut {NoStop}%
\bibitem [{\citenamefont {B\"{u}ttner}\ \emph {et~al.}(2011)\citenamefont
  {B\"{u}ttner}, \citenamefont {Liu}, \citenamefont {Tkachov}, \citenamefont
  {Novik}, \citenamefont {Br\"{u}ne}, \citenamefont {Buhmann}, \citenamefont
  {Hankiewicz}, \citenamefont {Recher}, \citenamefont {Trauzettel},
  \citenamefont {Zhang},\ and\ \citenamefont {Molenkamp}}]{Buttner2011}%
  \BibitemOpen
  \bibfield  {author} {\bibinfo {author} {\bibfnamefont {B.}~\bibnamefont
  {B\"{u}ttner}}, \bibinfo {author} {\bibfnamefont {C.~X.}\ \bibnamefont
  {Liu}}, \bibinfo {author} {\bibfnamefont {G.}~\bibnamefont {Tkachov}},
  \bibinfo {author} {\bibfnamefont {E.~G.}\ \bibnamefont {Novik}}, \bibinfo
  {author} {\bibfnamefont {C.}~\bibnamefont {Br\"{u}ne}}, \bibinfo {author}
  {\bibfnamefont {H.}~\bibnamefont {Buhmann}}, \bibinfo {author} {\bibfnamefont
  {E.~M.}\ \bibnamefont {Hankiewicz}}, \bibinfo {author} {\bibfnamefont
  {P.}~\bibnamefont {Recher}}, \bibinfo {author} {\bibfnamefont
  {B.}~\bibnamefont {Trauzettel}}, \bibinfo {author} {\bibfnamefont {S.~C.}\
  \bibnamefont {Zhang}}, \ and\ \bibinfo {author} {\bibfnamefont {L.~W.}\
  \bibnamefont {Molenkamp}},\ }\href@noop {} {\bibfield  {journal} {\bibinfo
  {journal} {Nature Physics}\ }\textbf {\bibinfo {volume} {7}},\ \bibinfo
  {pages} {418} (\bibinfo {year} {2011})}\BibitemShut {NoStop}%
\bibitem [{\citenamefont {Jiang}\ \emph {et~al.}(2011)\citenamefont {Jiang},
  \citenamefont {Kane},\ and\ \citenamefont {Preskill}}]{Jiang2011}%
  \BibitemOpen
  \bibfield  {author} {\bibinfo {author} {\bibfnamefont {L.}~\bibnamefont
  {Jiang}}, \bibinfo {author} {\bibfnamefont {C.}~\bibnamefont {Kane}}, \ and\
  \bibinfo {author} {\bibfnamefont {J.}~\bibnamefont {Preskill}},\ }\href@noop
  {} {\bibfield  {journal} {\bibinfo  {journal} {Phys. Rev. Lett.}\ }\textbf
  {\bibinfo {volume} {106}},\ \bibinfo {pages} {130504} (\bibinfo {year}
  {2011})}\BibitemShut {NoStop}%
\bibitem [{\citenamefont {Rojas-S\'anchez}\ \emph {et~al.}(2016)\citenamefont
  {Rojas-S\'anchez}, \citenamefont {Oyarz\'un}, \citenamefont {Fu},
  \citenamefont {Marty}, \citenamefont {Vergnaud}, \citenamefont {Gambarelli},
  \citenamefont {Vila}, \citenamefont {Jamet}, \citenamefont {Ohtsubo},
  \citenamefont {Taleb-Ibrahimi}, \citenamefont {Le~F\`evre}, \citenamefont
  {Bertran}, \citenamefont {Reyren}, \citenamefont {George},\ and\
  \citenamefont {Fert}}]{Rojas2016}%
  \BibitemOpen
  \bibfield  {author} {\bibinfo {author} {\bibfnamefont {J.-C.}\ \bibnamefont
  {Rojas-S\'anchez}}, \bibinfo {author} {\bibfnamefont {S.}~\bibnamefont
  {Oyarz\'un}}, \bibinfo {author} {\bibfnamefont {Y.}~\bibnamefont {Fu}},
  \bibinfo {author} {\bibfnamefont {A.}~\bibnamefont {Marty}}, \bibinfo
  {author} {\bibfnamefont {C.}~\bibnamefont {Vergnaud}}, \bibinfo {author}
  {\bibfnamefont {S.}~\bibnamefont {Gambarelli}}, \bibinfo {author}
  {\bibfnamefont {L.}~\bibnamefont {Vila}}, \bibinfo {author} {\bibfnamefont
  {M.}~\bibnamefont {Jamet}}, \bibinfo {author} {\bibfnamefont
  {Y.}~\bibnamefont {Ohtsubo}}, \bibinfo {author} {\bibfnamefont
  {A.}~\bibnamefont {Taleb-Ibrahimi}}, \bibinfo {author} {\bibfnamefont
  {P.}~\bibnamefont {Le~F\`evre}}, \bibinfo {author} {\bibfnamefont
  {F.}~\bibnamefont {Bertran}}, \bibinfo {author} {\bibfnamefont
  {N.}~\bibnamefont {Reyren}}, \bibinfo {author} {\bibfnamefont {J.-M.}\
  \bibnamefont {George}}, \ and\ \bibinfo {author} {\bibfnamefont
  {A.}~\bibnamefont {Fert}},\ }\href@noop {} {\bibfield  {journal} {\bibinfo
  {journal} {Phys. Rev. Lett.}\ }\textbf {\bibinfo {volume} {116}},\ \bibinfo
  {pages} {096602} (\bibinfo {year} {2016})}\BibitemShut {NoStop}%
\end{thebibliography}%

\onecolumngrid

\begin{LARGE}
\textbf{Supplementary materials: Revealing topological Dirac fermions at the surface of strained HgTe thin films via Quantum Hall transport spectroscopy.}\\
\end{LARGE}

In the main text of this letter, we have presented the results obtained on a first sample (called sample 1) fabricated from a 15 nm-thick HgTe layer surrounded by two Hg$_{0.3}$Cd$_{0.7}$Te barriers on a (100) CdTe substrate. These supplementary materials aim to complete the analysis by, first of all, adding information about the metallic behavior of the longitudinal resistance at zero magnetic field, the Landau level splitting as well as the Landau level broadening. To do so, experimental results from a second sample (called sample 2), fabricated from another 15 nm-thick HgTe layer, are presented and discussed. It is worth noting that the two samples have been made using the same growth and nanofabrication recipes. 
Finally, a model describing the hybridization origin of the Landau level splitting is also developed thus providing the justification of the observed integer quantization. 
With such contents, these supplementary materials aim to strengthen our experimental investigation of thin films of strained HgTe 3D topological insulators.

\maketitle 

\section{Longitudinal resistance $R_{xx}$ as a function of the Hall bar length}
\begin{figure}[h!]
\begin{center}
\includegraphics[scale=1]{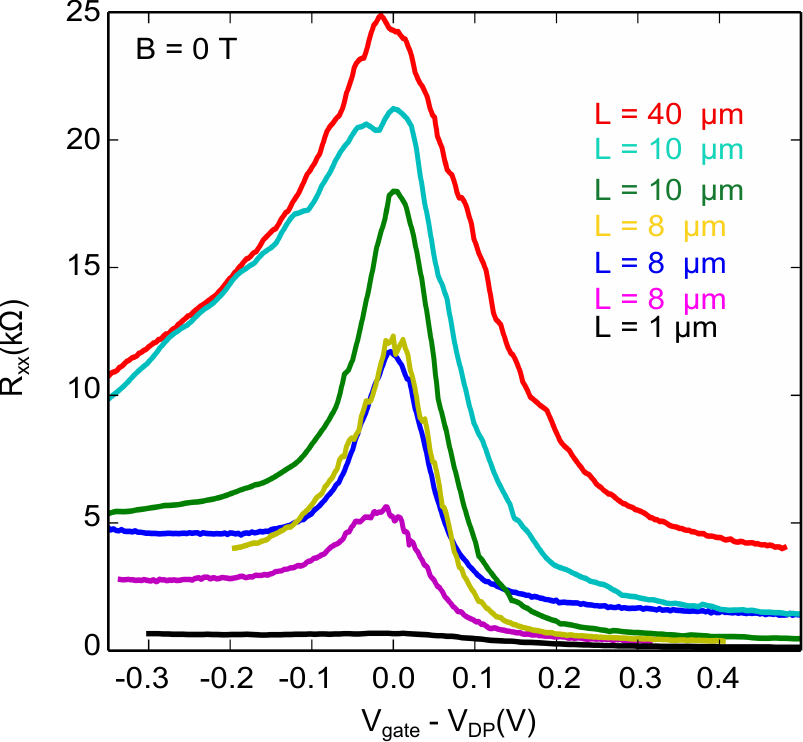} 
\caption{ Longitudinal resistance $R_{xx}$ as a function of $V_{gate}-V_{DP}$ at zero magnetic field for sample 2. $V_{DP}$ is defined for each curve as the voltage $V_{gate}$ corresponding to the maximum $R_{xx}$ and slightly varies from one Hall bar to another. The Hall bar length varies from 1 $\mu$m to 40 $\mu$m.  } \label{fig1SM}   
\end{center} \end{figure}

The gate dependence of $R_{xx}$ was measured at zero magnetic field for several Hall bars with lengths $L$ ranging from 1 to 40 $\mu$m keeping the aspect ratio constant. All the Hall bars considered in this section were fabricated from sample 2. Similar observations were performed on sample 1. Figure \ref{fig1SM} displays R$_{xx}$ as a function of the gate voltage for B=0 T and for T lower than 100 mK. For all the bars, $R_{xx}$ evolves the same way as a function of the gate voltage $V_{gate}$ and exhibits a maxima at the Dirac point (DP) for $V_{gate}=V_{DP}$. Resistance maxima are to first approximation increasing with $L$ from 1 k$\Omega$ to 25 k$\Omega$. This witnesses a non-ballistic transport suggesting a metallic behaviour of the surface states for which the resistance value is governed by scattering disorder potential. These observations are inconsistent with quantum spin Hall effect expected for thin HgTe topological quantum wells $[1]$. It is worth noting some divergences in the resistance of Hall bars of same dimensions.

\newpage

\section{Experimental evidence of the Landau level splitting in the hole-side}

In the main text, Hall bar magnetoconductance measurements were presented for magnetic fields up to 3 T where quantization between electron- and hole-side differs. All integer filling factors were observed in the electron-side whereas only odd ones for holes. Additional measurements have been performed for magnetic field values $B$ up to 5.5 T on sample 2.

\begin{figure}[!h]
\begin{center}
\includegraphics[scale=1]{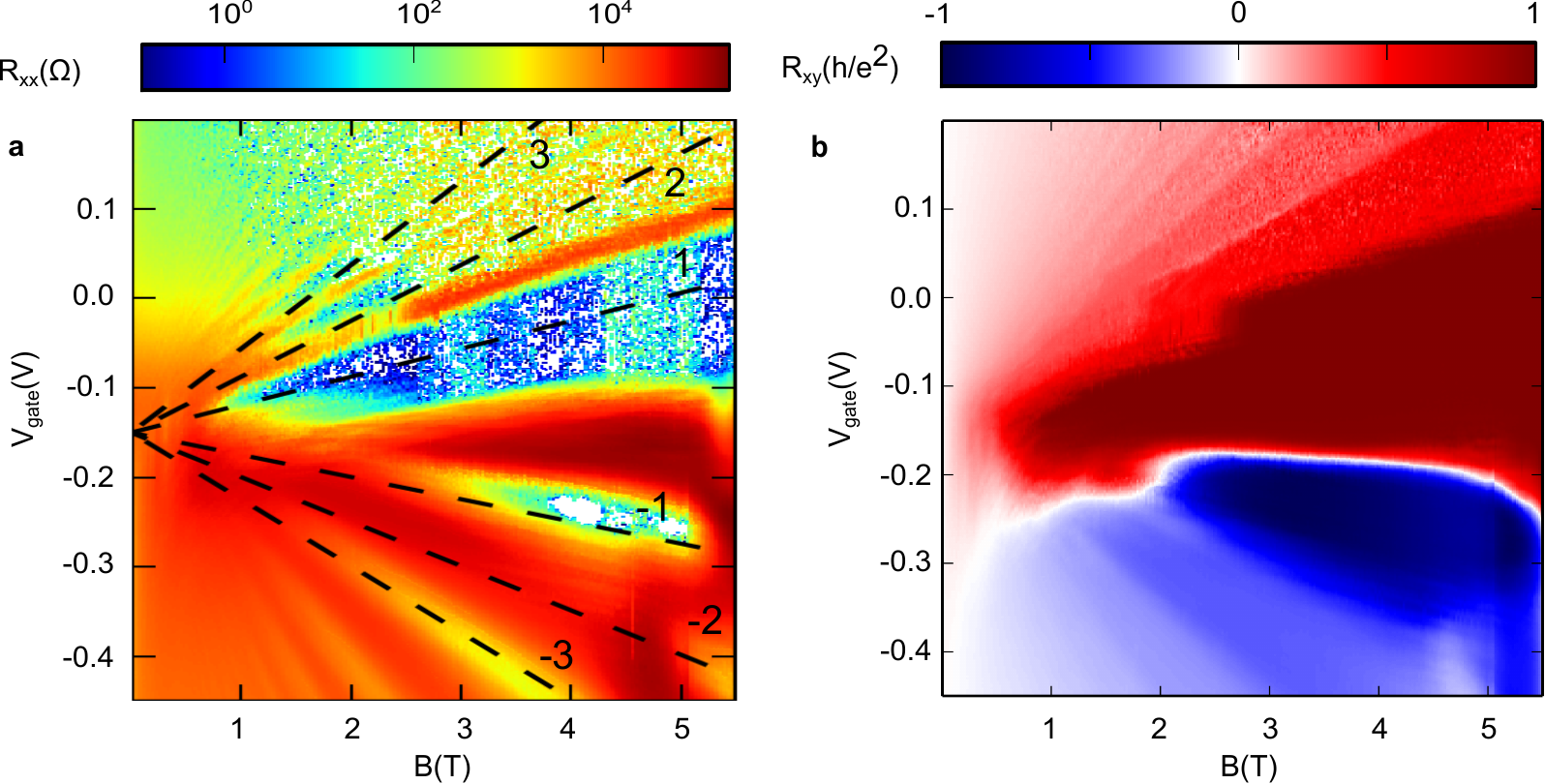} 
\caption{(a) $R_{xx}$ as a function of $V_{gate}$ and the perpendicular magnetic field $B$. The black dashed lines represent the expected position of the  minima of $R_{xx}$ from the density evolution with $V_{gate}$ extracted at low magnetic field (see Fig. 1).  (b) Hall resistance $R_{xy}$ (in unit of $\frac{h}{e^2}$) as a function of $V_{gate}$ and $B$. These data are extracted from sample 2.} \label{fig2SM}   
\end{center} \end{figure}

\begin{figure}[!h]
\begin{center}
\includegraphics[scale=1]{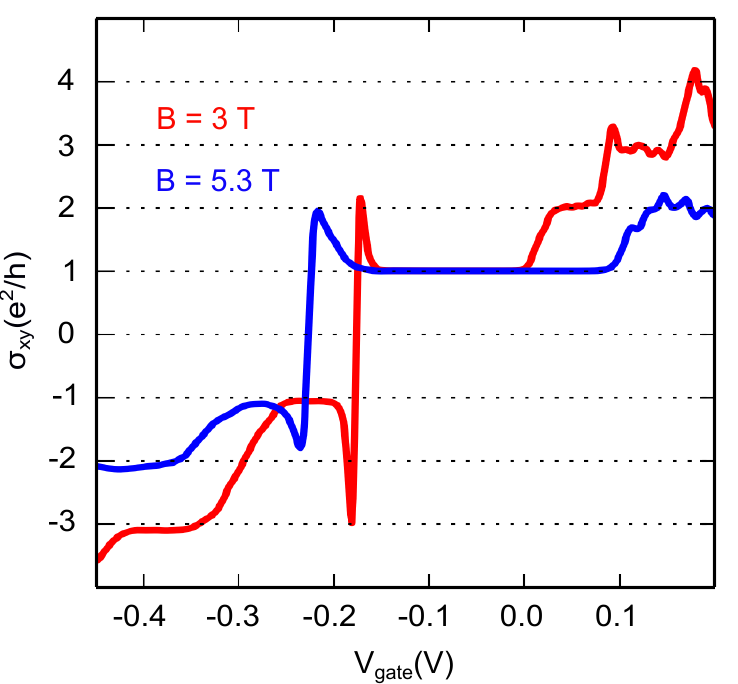} 
\caption{Appearance of even filling factor on the hole-side for high magnetic fields. Hall conductance $\sigma_{xy}$ as a function of $V_{gate}$ at $B=3$ T (red) and  $B=5.3$ T (blue) for sample 2.} \label{fig3SM}   \end{center}
\end{figure}

In this section, the Hall bar of interest is characterized by a length L = 8 $\mu$m and a width W = 2 $\mu$m. Supplementary Fig. \ref{fig2SM}  displays the longitudinal resistance $R_{xx}$ as well as the Hall resistance $R_{xy}$ as a function of B and $V_{gate}$ for T = 100 mK. Quantum Hall effect is also well-defined for this sample with vanishing minima of $R_{xx}$ directly associated with plateaus of $R_{xy}$. Using the same analysis than in the main text of the letter, the black dashed lines represent the expected filling factors from the low-magnetic field Hall measurement. These two mappings unambiguously evidence the appearance of the $\nu = -2$ plateau for B $\geq 5$ T. This is emphasized on Supplementary Fig. \ref{fig3SM} where two traces of  $\sigma _{xy}$ for both B=3 T and B=5.3 T are displayed. A very clear and well-defined $\nu = -2$ plateau exists for B=5.3 T whereas for B=3 T the quantization in the hole side jumps from -1 to -3.\\

The observed difference in the required magnetic field to resolve the Landau level splitting in the hole- and electron-side is explained by the presence of the  $\Gamma_{8,HH}$ bulk band. It results in the enlargement of the Landau level (LL) width in the hole-side and, in comparison with the electron-side, it then requires larger magnetic fields to resolve the even filling factors. 

\newpage

\section{R$_{xx}$ temperature dependance}
\begin{figure}[!h]
\begin{center}
\includegraphics[scale=1]{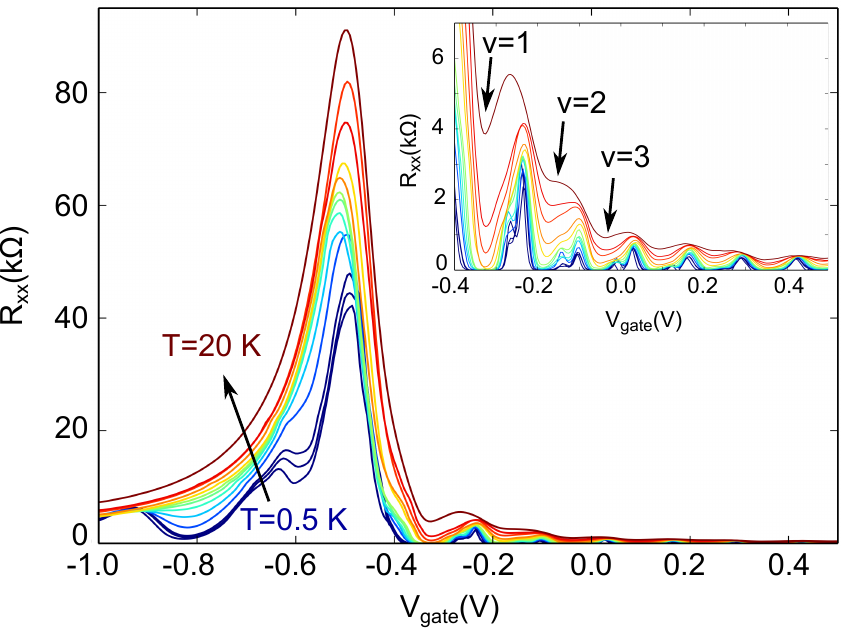} 
\end{center}
\caption{ $R_{xx}$ as a function of $V_{gate}$ for temperatures $T$ ranging from 0.5 K to 20 K for sample 1. The inset is a zoom on the electron-side. The thermally activated behavior of $R_{xx}$ minima is evidenced.}
\label{fig5SM}
\end{figure}

Quantum Hall effect measurements have been performed for temperatures going from the mK range to several Kelvins. In the main text, such analysis is used to extract the activation gap energy $\Delta E$ for odd and even filling factors presented in Fig. 4. Supplementary Fig. \ref{fig5SM} displays typical R$_{xx}$ of sample 1 as a function of the gate voltage for the different temperatures. The energy gap $\Delta E$ associated to each filling factors are evidenced as R$_{xx}$ minima. As expected, these resistance minima are increasing with temperature. One can notice that they are still visible at 20 K in the electron-side while they disappear around T= 4 K in the hole-side. Such difference is interpreted as the impact of the $\Gamma_{8,HH}$ bulk band on the hole-side. Such thermally activated behavior in the electron-side is fitted using Arrhenius law and evidences the Dirac nature of the charge carriers.

\newpage

\newpage

\section{Landau Level Broadening}

In the main text, Landau level broadening $\Gamma$ estimation is used to analyse the conditions of appearance of the splitting for both electron- and hole-sides.  As explained in $[2]$ and $[3]$, the analysis of the Shubnikov-de Haas (SdH) oscillation amplitude as a function of the magnetic field allows to obtain values of LL broadening through the determination of the quantum scattering lifetime $\tau_q$: 

\begin{equation}
\frac{\Delta R_{xx}}{4R_0} =\gamma_{th} \exp(-\frac{\pi}{w_c\tau_q})
\label{eq_broadening}
\end{equation}

where $\Delta R_{xx}$ is the amplitude of the SdH oscillations, $R_0$ is the resistance background of the oscillations, $\gamma_{th}$  describes the temperature effect on the LL broadening, and $w_c = \frac{eB}{m^*}$ is the cyclotron frequency. The cyclotron effective mass $m^*$ value is prior determined from the analysis of the SdH oscillation amplitude as a function of the temperature $[4]$ as explained in the following subsection.

\subsection{Estimation of the carrier cyclotron effective mass}
The cyclotron effective mass can be extracted from the temperature dependence of the SdH oscillation analysis
using the Lifshits-Kosevich (LK) formula:

\begin{equation}
\frac{\Delta R(B, T)}{\Delta R(B, T_{ref})} = \frac{T \sinh{(2 \pi^2 k_B T_{ref} m^*/\hbar e B})}{T_{ref} \sinh{(2 \pi ^2 k_B T m^*/\hbar e B})}
\label{eq_mstar}\end{equation}
with $k_B$ the Boltzmann constant.

\begin{figure}[!h] \begin{center}
\includegraphics[scale=1]{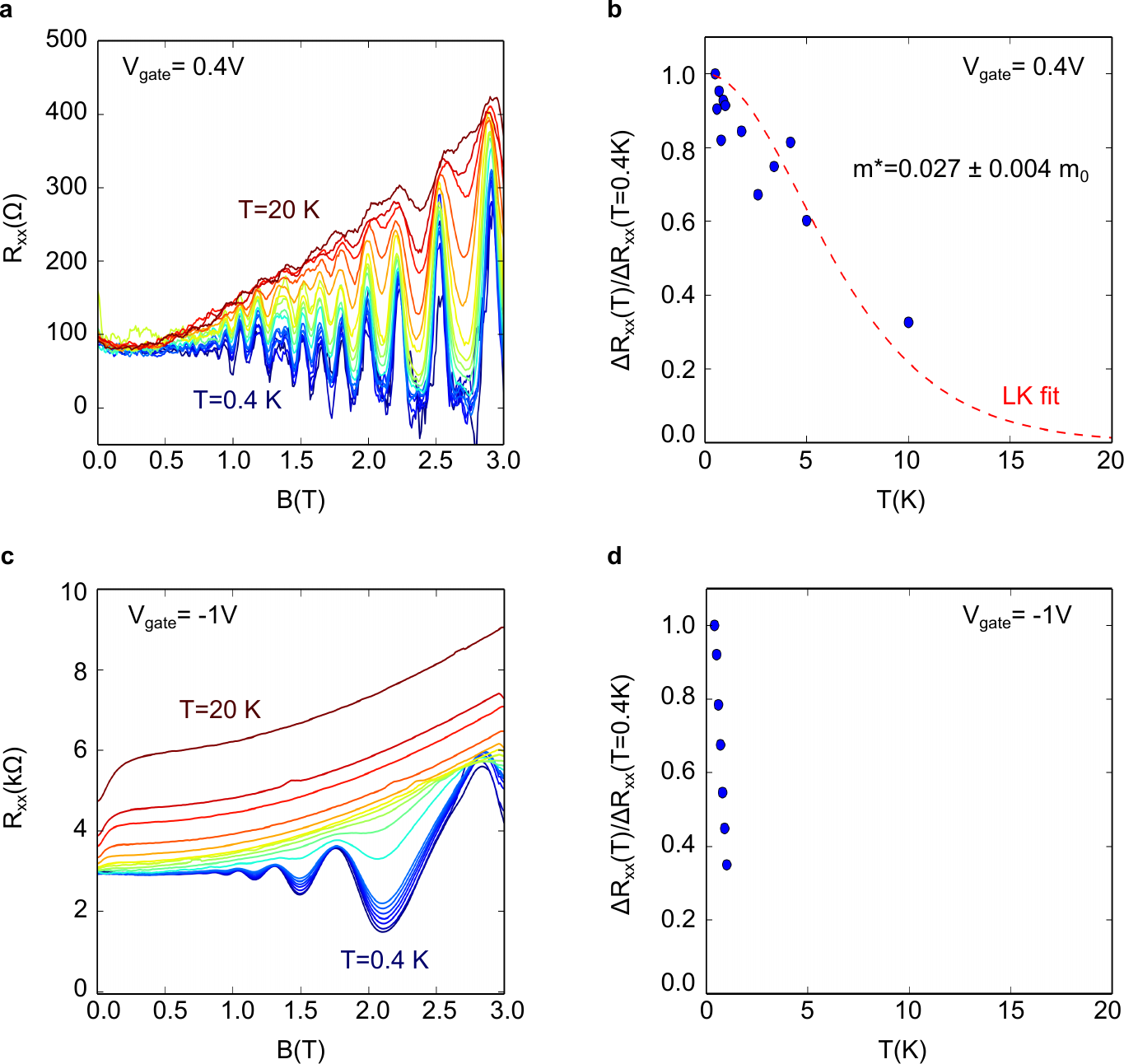} 
\caption{ SdH oscillations of the longitudinal resistance $R_{xx}$ of sample 1 for $V_{gate}=0.4$ V (a) and for $V_{gate}= -1$ V (c) for temperatures ranging from 0.4 K to 20 K. (b) The corresponding Lifshits-Kosevich (LK) fit for the extraction of the cyclotron effective mass $m^*$ on the electron-side. (d) Amplitude variations of the SdH oscillations on the hole-side.  } \label{m_star} 
\end{center} \end{figure}

Figure \ref{m_star}(a) displays $R_{xx}$ as a function of  $B$ for $V_{gate}=0.4$ V. The temperature $T$ is increased from 0.4 K to 20 K with 0.1 K steps below 1 K, 0.8 K steps between 1 and 5 K and finally 5 K steps between 5 and 20 K. The oscillation amplitude is decreasing with $T$ and the oscillations become hardly distinguishable at 20 K. The determination of the carrier cyclotron effective mass $m^*$  is thus possible using the data from Fig. \ref{m_star}(a) and equation \ref{eq_mstar} with $T_{ref}$ = 0.4 K. Note that amplitude variations of the oscillations have been studied for $B$ close to 1 T to avoid spin splitting and only consider the surface state carrier intrinsic properties.

 The curve of $\Delta R_{xx}(T)/\Delta R_{xx}(T=0.4 K)$  (see Fig. \ref{m_star}(b))  is fitted using the red dashed line demonstrating $m^*$ value of about $0.027 \pm 0.004$ $m_0$.  However, such fit is not so accurate. This is mainly due to the lack of data at larger temperatures but as can be seen in Fig. \ref{m_star}(a), oscillations for T larger than 10 K are difficult to resolve for small $B$. Due to this issue, determination of $m^*$ is difficult in our structures.

Similar study, but in the hole-side, is presented in Fig.\ref{m_star}(c) and (d). In this regime, the SdH oscillations disappear very quickly with temperature and are no more distinguishable above 3.4 K. This behavior is attributed to the influence of the $\Gamma_{8,HH}$ bulk band. To properly estimate  $m^*$ in this regime, we consider only $T$ from 0.4 to 1 K, range of $T$ for which the oscillations are the best defined. 
Amplitude dependence of SdH oscillations is reported in Fig.\ref{m_star}(d) and considerably differs  from what has just been reported for electrons. No saturation is visible at the lower temperatures, but on the contrary a striking increase is noticeable. Such a behavior can not be fitted by a standard metallic LK model. The coupling with bulk $\Gamma_{8,HH}$ states in the hole regime is expected to be at the origin of this particular feature. 


Nevertheless, it is possible to have an idea of the carrier cyclotron effective mass  in our HgTe structures by using analogy with graphene. 
Due to a linear energy spectrum, one expects a zero effective mass for Dirac fermions. However, as already reported for graphene $[5]$, Dirac fermions rest mass is zero but not the cyclotron mass. Note that SdH oscillation amplitude dependence with temperature gives directly access to this cyclotron mass  $[4]$.  However, standard definition for $m^*$ relies on energy second derivative and is only adapted for materials defined with parabolic dispersion. 
An alternative definition has been developed for materials with linear dispersion law $[6]$: 

\begin{equation}
m^* = \frac{p}{\frac{\delta E}{\delta p}}= \frac{\hbar k}{v_F}=\frac{\hbar \sqrt{\pi n}}{v_F}
\label{eq_mstar_dirac}
\end{equation}
This relation directly links  the cyclotron effective mass $m^*$  and the carrier density $n$. 

\begin{figure}[!h] \begin{center}
\includegraphics[scale=1]{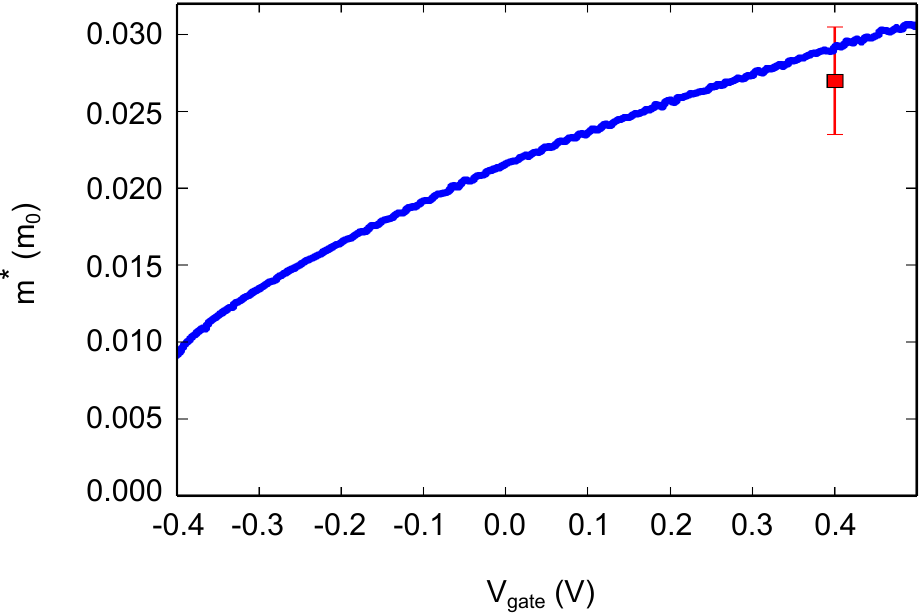} 
\caption{Carrier cyclotron effective mass for particles with linear energy dispersion. Evolution of $m^*$ as a function of $V_{gate}$ for a Dirac cone (blue dotted line). Estimation of $m^*$ using SdH oscillations has been performed for $V_{gate}=0.4$ V and is represented by a red square. Error bar corresponds to the LK fit error.} \label{m_star_sqrt_n} 
\end{center} \end{figure}

Fig. \ref{m_star_sqrt_n} displays the evolution of $m^*$, expected from the density  variation (see Fig. 1(c) in main text), as a function of $V_{gate}$. Note the red square which represents the effective mass estimated just before. These two methods appear to give consistent results. 
 Therefore, we will use relation $\ref{eq_mstar_dirac}$ to determine $m^*$ thereafter.

\newpage

\subsection{Estimation of the LL broadening}

Due to a non negligible bulk contribution in the hole-side, the LL broadening estimation method of equation \ref{eq_broadening} has only been performed in the electron-side where amplitude of SdH oscillations only depends on the surface state properties. 
$\tau_q$  is then directly related to $\Gamma$ through $\Gamma=\frac{\hbar}{2\tau_q}$.

\begin{figure}[!h]\begin{center}
\includegraphics[scale=1]{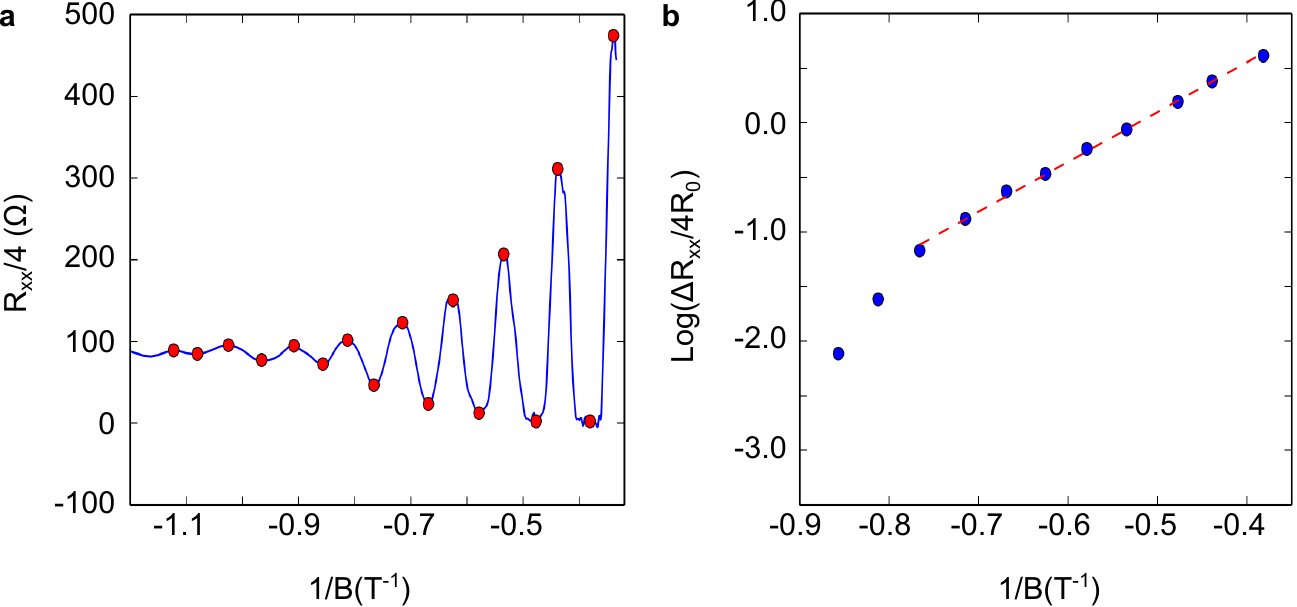} 
\caption{Estimation of the Landau level broadening for $V_{gate}=0$ V. (a) Longitudinal resistance $R_{xx}$ as a function of the inverse of the magnetic field for $V_{gate}=0$ ~V and $T=0.4$ K for sample 1. The red dots represent maxima and minima defining the envelope function. The difference between maxima and minima gives $\Delta R_{xx}$. (b) $\Delta R_{xx}$ normalized by the background $R_0$ as a function of the inverse of the magnetic field. Blue dots represent the data and the red dashed line is a fit using relation \ref{eq_broadening}.} \label{broadening_fit}
\end{center} \end{figure}

Figure \ref{broadening_fit}(a) presents the evolution of $R_{xx}$ as a function of the inverse of the magnetic field $1/B$ for $V_{gate}=0$ V at $T=0.4$ K.  The red dots represent both the maxima and the minima which define an envelope function. $\Delta R_{xx}$ is extracted from the difference between successive maxima and minima and $R_0$ results from the mean value of these extrema. 
Figure \ref{broadening_fit}(b) displays the evolution of $\Delta R_{xx}/R_0$ as a function of $1/B$ in a logarithmic scale. The red dashed line represents the linear fit to the data using equation \ref{eq_broadening}. The extracted value of $\tau_q$ provides direct access to the broadening value here equal to 3.91 $\pm$ 0.04 meV.

\begin{figure}[!h]
\begin{center}
\includegraphics[scale=1]{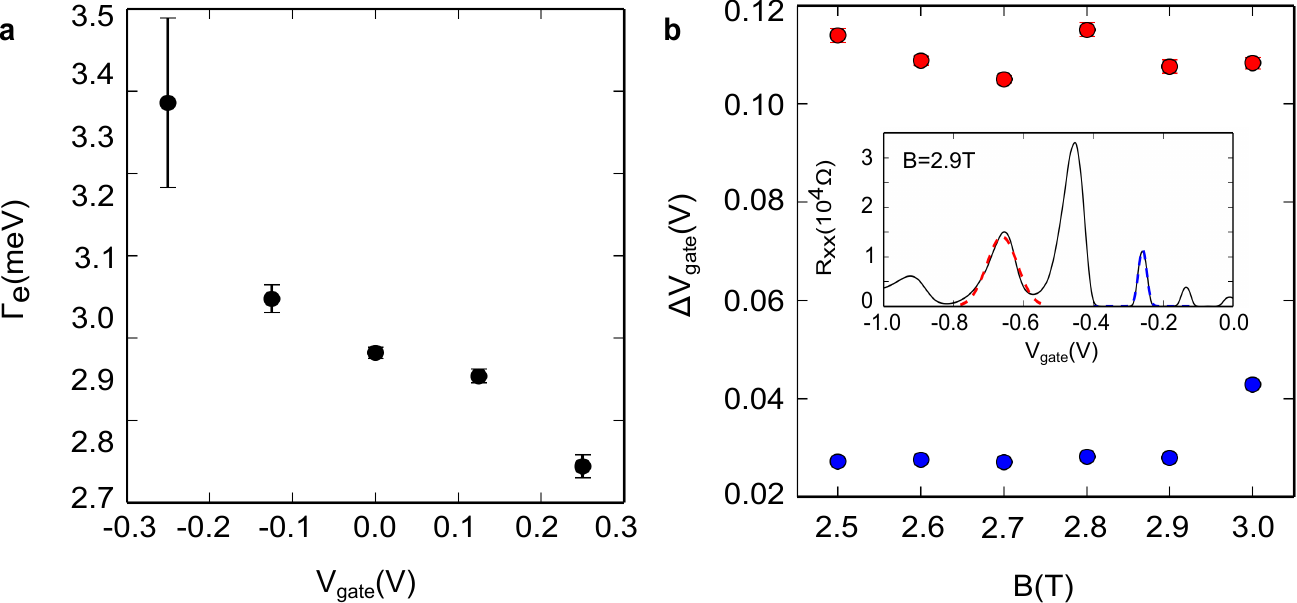} 
\end{center}
\caption{Estimation of the Landau level broadening on the electron- and the hole-sides. (a) Determination of the LL energy broadening $\Gamma_e$ as a function of V$_{gate}$ in the electron-side for sample 1. (b) Estimation of the energy broadening $\Gamma_h$ on the hole-side using comparison between the full width at half maximum $\Delta V_{gate}$ of the $R_{xx}$ peaks for the $\text{N}=1$ (blue) and $\text{N}=-1$ (red) peaks (shown in the inset) as a function of $B$. } \label{fig6SM}   
\end{figure}

The same analysis has been performed for several values of $V_{gate}$. 
Figure~\ref{fig6SM}(a) displays the extracted values of electron-side broadening $\Gamma_e$. Note that the broadening is decreasing with $V_{gate}$ and thus with density $n$. This behavior is due to the variations of $m^*$ with $n$ and is consistent with $R_{xx}$ mapping of Fig. 2(a) (see main text) Indeed, we can see that the closer are the LLs from the Dirac point, the more important is the magnetic field value at which the splitting appears. 
This is especially visible with the $\nu=2$ plateau appearing for $|B| \geq 1.5$ T while $\nu=4$ shows up at larger gate voltage value for $|B| \geq 1.0$ T. 

To have an estimation of the broadening in the hole-side, we have fitted $N=1$ and $N=-1$ LL peaks with a Gaussian distribution to extract the full width at half maximum (see inset of Fig. \ref{fig6SM}(b)). Figure \ref{fig6SM}(b) demonstrates the difference through the $\Delta V_{gate}$ values between hole and electron peaks. With a ratio of about 3.7, the hole broadening $\Gamma_h$ has been estimated to be larger than 11 meV. More precise determination of $\Gamma_h$ is complicated as it also depends  on the gate voltage. \\

\newpage

\newpage

\section{Landau level splitting appearance: discussion}

\begin{center}
\begin{figure}[h!]
\includegraphics[scale=1]{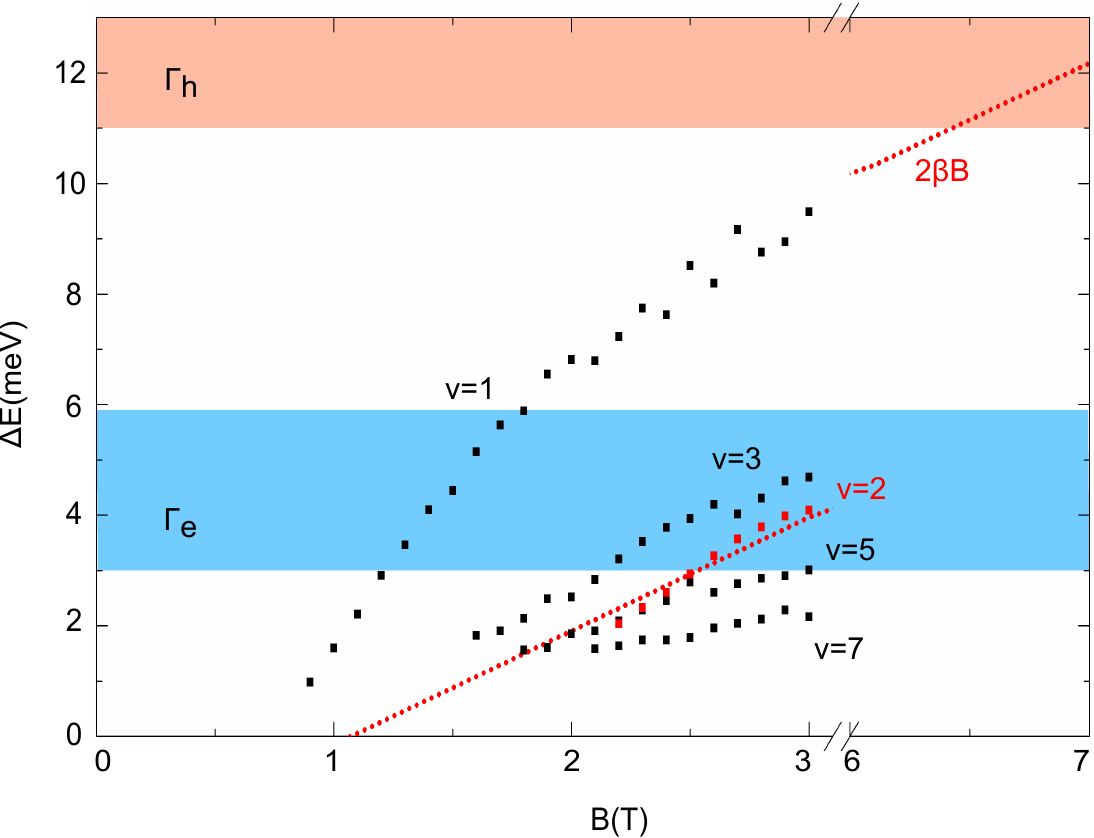} 
  \caption{Final fan diagram with the odd  $\nu$=1, 3, 5, 7 (black squares), and even $\nu$=2  (red squares) filling factors and the "effective Zeeman splitting" energy (red dotted line). $\Gamma_h$ (respectively $\Gamma_e$)  range is represented by a red (blue) color band. Note that the energy ranges considered for broadening values take into account variations with gate voltage.}\label{fig7SM} 
 \end{figure}\end{center}

Supplementary Fig.~\ref{fig7SM} summarizes on the same diagram all the relevant energy scales discussed in the letter. All the energy gaps $\Delta$E of both odd and even filling factors for different magnetic fields discussed in Fig. 4 (see main text) are displayed as squares. The color bands correspond to the Landau level broadening $\Gamma_e$ and $\Gamma_h$ extracted in the previous section. If the energy gaps are smaller than Landau level broadening, they can not be resolved by the quantum Hall transport spectroscopy. This diagram suggests the appearance of even filing factors in the electron side for $B_e \geq 2.4$ T and in the hole side for $B_h \geq 6.3$ T. Such values are consistent in term of ratio with experimental data for which $B_e$ and $B_h$ were estimated around 1.5 T and 4.5 T respectively. 

\section{Model and Origin of the Landau Level Splitting} 

\subsection{Model}
We consider a thin layer of topological insulator material, with two parallel surfaces indexed by $\tau^z=+1$ for the
top (T) and $\tau^z=-1$ for the bottom (B) surface. The effective Hamiltonian for the surface states reads
\begin{equation} 
H = v_{f}  ~\tau_{z}  \otimes ( \hat{z} \times \vec{\sigma} ) .( -i \hbar \vec{\nabla})  
+ \left( \frac{\Delta}{2}   +  \frac{\hbar^2 \nabla^2}{2M} \right) \tau_{x}    \otimes  \mathbb{I} 
+ V   \tau_{z} \otimes \mathbb{I} ,  
\label{eq:H}
\end{equation}
where $v_f$ is the Fermi velocity of the surface states and $\sigma_x , \sigma_y$ describes Pauli matrices acting on the spin Hilbert space while 
$\tau_\alpha$ acts on the T/B space. The first term describes the Dirac dispersion relation for the top and bottom surface states 
with opposite chirality  (the spin winds in opposite ways around the Fermi surface), while $\Delta/2 - (\hbar k)^2/2M$ is
a momentum dependent tunneling between the two surfaces, and $V$ the chemical
potential difference between the surfaces. 
The band structure corresponding to this model is symmetric with respect to $E=0$, and reads 
 \begin{equation}
E^2_\pm  ({\bf k})= 
\left( \frac{\Delta}{2} -  \frac{\hbar^2 k^{2}}{2M} \right)^2 + \left( V \pm \hbar v_f |k| \right)^2 . 
\end{equation}

\subsection{Landau spectrum}

We consider a magnetic field $B$ applied along $z$, perpendicular to the two surfaces. We
 obtain a spectrum with a chiral symmetry with respect to $E=0$. The
 Zeeman coupling is neglected in the Hamiltonian (\ref{eq:H}), so the effect of the magnetic field is purely orbital, and
 enters into the Hamiltonian by the minimal substitution $-i \hbar
 \vec{\nabla} \to  -i \hbar
 \vec{\nabla} - e \vec{A}$ where $\vec{A}$ is the electromagnetic
 vector potential. 
 In the gauge $(A_x = 0 , A_y = B x)$, we introduce the ladder operators 
 \begin{align}
 a &= \frac{1}{\sqrt{2}} ( q+i p ) 
 \ ;\  
 a^\dagger =  \frac{1}{\sqrt{2}} ( q-i p ) ,\\
 q &= \frac{X}{l_B} = \frac{x - k_y l_B^2}{l_B^2} 
 \ ;\ 
 p = -i l_B \partial_X , 
 \end{align}
 where we introduced the magnetic length 
  $ l_{B}^{2} = \hbar / eB $. 
 The Hamiltonian in the presence of a magnetic field can be rewritten as 
 \begin{equation}
H = 
- \eta ~\tau_z \otimes \left( a \sigma_- + a^\dagger \sigma_+ \right)
+ \left(  \frac{\Delta}{2}- \frac{\hbar \omega}{2} (2 a^\dagger a +1 )    \right) \tau_x \otimes \mathbb{I} 
+ V \tau_z \otimes \mathbb{I}
\end{equation}
where we introduced  $2\sigma_\pm = \sigma_x \pm i \sigma_y$ and 
\begin{equation}
\eta^2 = 2 \frac{ \hbar^2 v_f^2 }{l_B^2}  = 2 \hbar v_F^2 e B 
\quad ; \quad 
\hbar \omega = \frac{\hbar^2}{M l_{B}^{2}} = \frac{\hbar e B}{M}  . 
\end{equation}
We consider the basis $|N\rangle$ of quanta of the operators $a,a^\dagger$. 
In the basis 
$
| T , \uparrow , N \rangle , 
| T , \downarrow , N-1  \rangle , 
| B ,  \uparrow , N  \rangle , 
| B ,  \downarrow , N-1  \rangle , 
$
the reduced Hamiltonian reads for $N\ge 1$: 
\begin{equation}
H_n = 
\begin{pmatrix}
V , -\eta \sqrt{N} , \frac{\Delta}{2} - \frac{\hbar \omega}{2} ( 2 N +1)  , 0 \\
 -\eta \sqrt{N} , V , 0 ,   \frac{\Delta}{2} - \frac{\hbar \omega}{2} ( 2 N -1)  \\
 \frac{\Delta}{2} - \frac{\hbar \omega}{2} ( 2 N +1)   , 0 , - V , \eta \sqrt{N} \\
  0 ,  \frac{\Delta}{2} - \frac{\hbar \omega}{2} ( 2 N -1)   , \eta \sqrt{N} , -V 
\end{pmatrix}  
= 
\tau_x \otimes \left[  
\left( \frac{\Delta}{2} - N \hbar \omega   \right) \mathbb{I} + \frac{\hbar \omega}{2} \sigma_z \right]  
\\
- \eta \sqrt{N} ~   \tau_z \otimes \sigma_x 
+ V \tau_{z}  \otimes  \mathbb{I} . 
\label{eq:Hn}
\end{equation}
Its spectrum is symmetric with respect to $0$ (chiral symmetric) and the energies satisfy 
\begin{equation}\label{eq:spectrum-n-pos} 
E_{N, \pm}^{2} = 
V^{2} +  N  \eta^2  
+ \left( \frac{\hbar \omega }{ 2 } \right)^2 
+ \left(  \frac{\Delta }{2} - N \hbar\omega \right)^{2}
\pm 
\left( 
 N \eta^2   \left( 4 V^{2} +   (\hbar\omega)^2     \right) 
+ (\hbar\omega)^2  \left(   \frac{\Delta}{2} - N \hbar\omega \right)^{2} 
\right)^{\frac12} 
\textrm{ for } N\neq 0 , 
\end{equation}
which identifies exactly with the spectrum of $[7]$ with the notation  $B   =  \hbar^2 /2M $. 

In the case $N=0$, we have to consider the basis $|T,\uparrow , 0\rangle$
and $|B,\uparrow, 0\rangle$. These basis states are annihilated
by $a\sigma_-$ and by $a^\dagger\sigma_+$, and the effective
Hamiltonian is then:
\begin{eqnarray}
  H_0=V \tau_z \otimes \openone +\left(\frac {\hbar \omega -\Delta}
    2\right) \tau_x \otimes \openone,   
\end{eqnarray}
with eigenvalues $\pm \sqrt{V^2 +(\hbar \omega-\Delta)^2/4}$.

\subsection{Landau spectrum with $V=0$}

In the special case $V=0$, we can rewrite Eq.~(\ref{eq:spectrum-n-pos}) as 
\begin{subequations}
\begin{align}
\tilde{E}_{N, \pm}^{2}  &= 
  N  \eta^2  
+ \left( \frac{ \hbar\omega }{ 2 } \right)^2 
+ \left(  \frac{\Delta }{2} - N \hbar\omega   \right)^{2}
\pm\hbar \omega 
\left( 
 N \eta^2      
+   \left(  \frac{\Delta}{2} - N\hbar \omega  \right)^{2} 
\right)^{\frac12} 
=  \left( 
\epsilon_N   \pm  \frac{\hbar \omega }{ 2 }  
\right)^2
\\
\tilde{E}_{0}^{2}  &=  \left(\frac{\hbar \omega-\Delta}{2} \right)^2
\end{align}
\end{subequations}
where 
\begin{equation}
\epsilon_N = \left( 
 N \eta^2       
+   \left(   \frac{\Delta}{2} -  N \hbar\omega \right)^{2} . 
\right)^{\frac12}
\end{equation}
Hence the degeneracy of the Landau levels $\epsilon_{N}$ of the two relativistic surfaces corrected by the non-relativistic couplings is lifted by a 
splitting   $\pm \hbar \omega /2 $ ($\pm \beta$ in the notation of the main text).  The $N=0$ Landau level was initially not degenerate. 
Note that while this splitting is not of magnetic origin, it is linear in magnetic field, and can be 
viewed as an "effective Zeeman splitting" originating from the $k^2$ dependence of hybridization term between the Top and 
Bottom surface $[7]$.   Such Landau energy spectrum is represented in Supplementary Fig. \ref{fig8SM}.

\subsection{Nature of the "Landau Level Splitting"}

When $V=0$ and in the limit $\Delta \to 0, \omega = 0$ the reduced Hamiltonian (\ref{eq:Hn}) reads 
 \begin{equation}
H_N^{(0)} = 
- \eta \sqrt{N} ~  \tau_z \otimes \sigma_x . 
\label{eq:Hn0}
\end{equation}
The corresponding Laudau eigenstates are (in the gauge we have chosen): 
\begin{subequations}
\begin{align}
& \psi_{N,+}^{+} = 
\frac{1}{\sqrt{2}} \left(| T  , \uparrow , N  \rangle - | T , \downarrow , N-1 \rangle \right) , 
& \psi_{N,-}^{+} = 
\frac{1}{\sqrt{2}} \left( | B  , \uparrow , N  \rangle + | B , \downarrow , N-1 \rangle \right) 
& \textrm{ for  } \epsilon_N = \sqrt{N} \eta , 
\\
& \psi_{N,+}^{-} = 
\frac{1}{\sqrt{2}} \left( | B , \uparrow , N   \rangle - | B , \downarrow  , N-1\rangle \right) , 
& \psi_{N,-}^{-} = 
\frac{1}{\sqrt{2}} \left(| T  , \uparrow, N  \rangle + | T, \downarrow , N-1  \rangle \right) 
& \textrm{ for  } \epsilon_N = - \sqrt{N} \eta . 
\end{align}
\end{subequations}
 They are entirely localized in either the top (T) or bottom (B) surface,  and carry no net magnetization: 
\begin{equation}
\langle  \psi_{N,\pm}^{+} | S_{x} |   \psi_{N,\pm}^{+} \rangle = 
\langle  \psi_{N,\pm}^{+} | S_{y} |   \psi_{N,\pm}^{+} \rangle =
 \langle  \psi_{N,\pm}^{+} | S_{z} |   \psi_{N,\pm}^{+} \rangle = 
 0 . 
\end{equation}

 Let us now discuss the effect of various terms of the Hamiltonian (\ref{eq:Hn}), treated as perturbations of the relativistic Landau Hamiltonian (\ref{eq:Hn0}): 
\begin{itemize}
\item Chemical Potential Asymmetry $V$. The corresponding term is  
$+ V \tau_{z}  \otimes  \mathbb{I}$ which merely shifts the energies of the Landau levels of the top surface with respect to those of the bottom surface without 
modifying their nature. 

\item Top / Bottom Hybridization: the corresponding term is 
$\left(  \frac{\Delta}{2} - N \hbar\omega  \right) \tau_x \otimes  \mathbb{I} $. This operator will shift the energies of the Landau levels and delocalize the eigenstates 
on both top and bottom surfaces but without any splitting between the two degenerate eigenstates. This can be inferred from the matrix elements  
\begin{align}
 \tau_x \otimes  \mathbb{I}  |   \psi_{N,s}^{\pm} \rangle  = |   \psi_{N,s}^{\mp} \rangle 
 \Rightarrow 
 \langle  \psi_{N,+}^{+ } |  \tau_x \otimes  \mathbb{I}  |   \psi_{N,-}^{+} \rangle = 0 . 
\end{align}

\item Splitting. 
The most striking consequence of the coupling between the two surface states arises from the $k^{2}$ term in (\ref{eq:H}) 
and leads in the reduced Hamiltonian (\ref{eq:Hn}) to a term 
\begin{equation}
H_{Splitting} = 
 \frac{\hbar \omega}{2} ~  \tau_x  \otimes \sigma_z  , 
\end{equation}
which satisfies 
\begin{equation}
H_{Splitting}  |   \psi_{N,+}^{+} \rangle = |   \psi_{N, - }^{+} \rangle . 
\end{equation}
Indeed, while the two eigenstates $|   \psi_{N,\pm}^{+} \rangle $ carry no magnetization, the spin and orbital degree of freedom (eigenstates $N$) 
are tightly bound together in their structure.  This is illustrated by the absence of magnetization along $x$. It is this relation between spin 
and orbital degree of freedom which is broken by the perturbation $H_{Splitting} $, whose eigenstates are  symmetric (S) or 
antisymmetric (A) between the two surfaces and read respectively for the  positive and negative eigenvalues: 
\begin{subequations}
\begin{align}\label{eq:pos-eigen}
& | \uparrow , S   \rangle = 
\frac{1}{\sqrt{2}} \left( | T , \uparrow, N  \rangle + | B ,  \uparrow , N \rangle \right) , 
& 
   | \downarrow , A  \rangle  = 
\frac{1}{\sqrt{2}} \left(| T ,  \downarrow , N-1  \rangle - | B ,  \downarrow , N-1  \rangle \right)  ,  
\\
%
& | \uparrow , A  \rangle = 
\frac{1}{\sqrt{2}} \left( | T , \uparrow , N \rangle - | B ,  \uparrow , N \rangle \right)  ,  
&| \uparrow , S   \rangle = 
\frac{1}{\sqrt{2}} \left(| T ,  \downarrow  , N-1 \rangle + | B ,  \downarrow , N-1   \rangle \right)  . 
\end{align}
\end{subequations}

Note that these eigenstates carry a magnetization along $z$: the splitting of the Landau level eigenstates 
will be associated with the appearance of finite magnetization along $z$ of the eigenstates, linear in magnetic field (in $\omega$). 
\end{itemize}
%

\begin{figure}[!h]
\begin{center}
\includegraphics[scale=1]{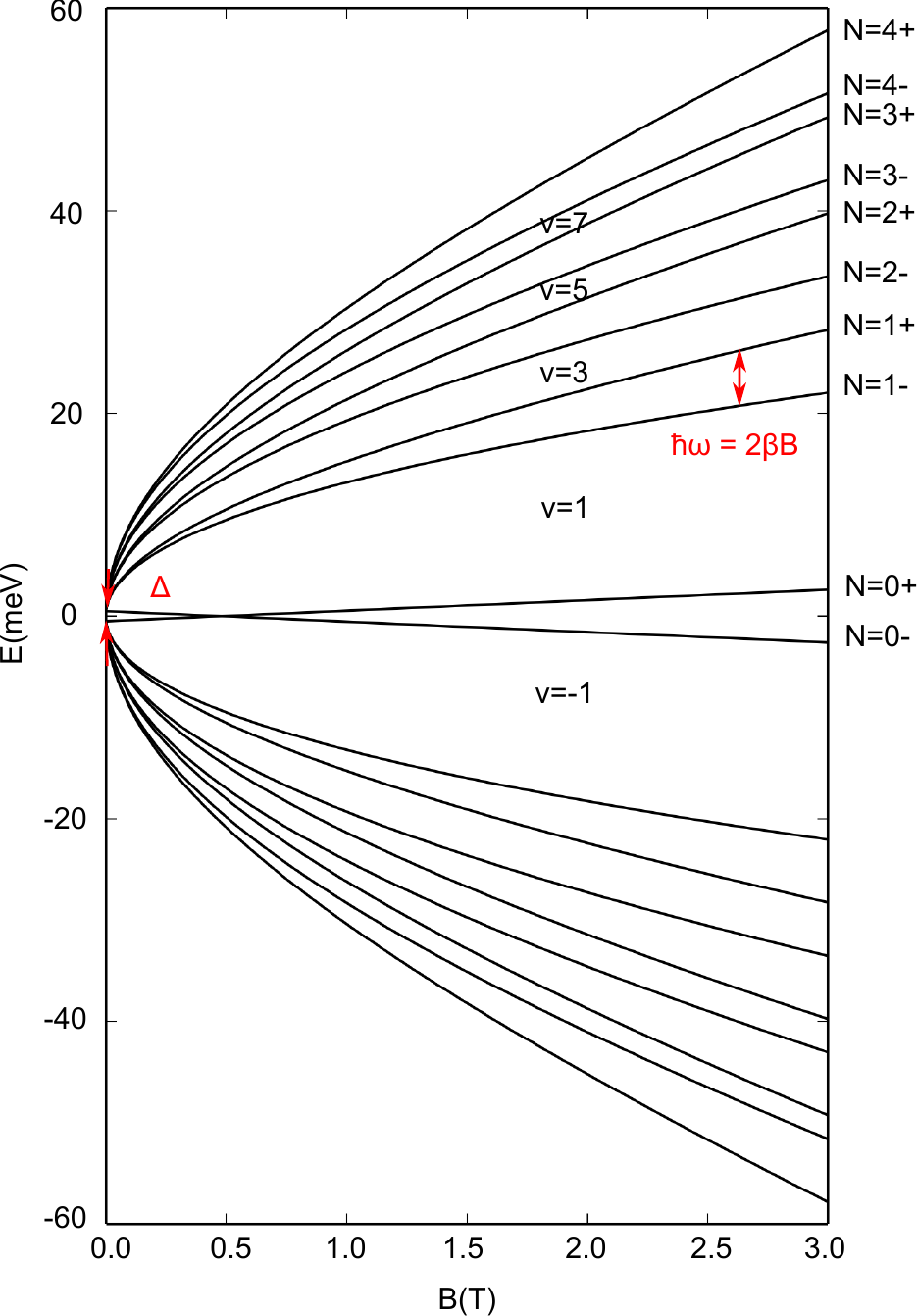} 
\caption{ Energy spectrum as a function of magnetic field for LL index $N$ ranging from 0 to 4 on both electron- and hole-sides. LL splitting as well as hybridization gap $\Delta$ are evidenced. $\Delta$ has been set to 2 meV, $2 \beta$ to 2.07 meV/T and $v_f$ to $\approx 4. 10^5$ m.s$^{-1}$.  }
\label{fig8SM}
\end{center} \end{figure}

\newpage
Bibliography:\\
$[1]$ K\"{o}nig, M. \textit{et al.} Quantum spin Hall insulator state in HgTe quantum wells. \textit{Science}, \textbf{318}, 766-770 (2007).\\
$[2]$ Hong, X., Zou, K. \& Zhu, J. Quantum scattering time and its implications on scattering sources in graphene. \textit{Phys. Rev. B}, \textbf{80}, 241415(R) (2009).\\
$[3]$ Young, A.F. \textit{et al.} Spin and valley quantum Hall ferromagnetism in graphene. \textit{Nature Physics}, \textbf{8}, 550-556 (2012).\\
$[4]$ Fuchs, J.-N. Dirac fermions in graphene and analogues: magnetic field and topological properties. \textit{Habilitation à diriger des recherches} (2013). \\
$[5]$ Novoselov, K. \textit{et al.} Two-dimensional gas of massless Dirac fermions in graphene. \textit{Nature}, \textbf{438}, 197-200 (2005).\\
$[6]$ Ariel, V. \& Natan, A. Electron Effective Mass in Graphene. \textit{arXiv}, 1206.6100 (2012). \\
$[7]$ Zhang, S., Lu, H. \& Shen, S. Edge states and integer quantum
Hall effect in topological insulator thin films. \textit{Scientific Reports}, \textbf{5}, 13277 (2015).

\end{document}